\documentclass[float,epsfig,usenatbib]{mn2e}

\usepackage{graphicx}
\usepackage{appendix}
\usepackage{psfrag}
\usepackage{amsmath,amssymb}
\usepackage{threeparttable}
\usepackage{float}
\usepackage{subfigure}
\usepackage{afterpage}

\usepackage{graphicx}
\usepackage{txfonts}
\usepackage{natbib}
\usepackage{lscape}
\usepackage{longtable}

\begin{document}

\def\deg{$^{\circ}$}
\newcommand{\eg}{{\it e.g.}}
\newcommand{\ie}{{\it i.e.}}
\newcommand{\minusone}{$^{-1}$}
\newcommand{\kms}{km~s$^{-1}$}
\newcommand{\kmsm}{km~s$^{-1}$~Mpc$^{-1}$}
\newcommand{\Ha}{$\rm H\alpha$}
\newcommand{\Hb}{$\rm H\beta$}
\newcommand{\hi}{H{\sc i}}
\newcommand{\hii}{H{\sc ii}}
\newcommand{\htwo}{H$_2$}
\newcommand{\nii}{\ion{N}{2}}
\newcommand{\rband}{{\em r}-band}
\newcommand{\iband}{{\em i}-band}
\newcommand{\zband}{{\em z}-band}
\newcommand{\rd}{$r_{\rm d}$}
\newcommand{\whi}{$W_{50}$}
\newcommand{\ds}{$\Delta s$}
\newcommand{\x}{$\times$}
\newcommand{\about}{$\sim$}
\newcommand{\Msun}{M$_\odot$}
\newcommand{\Lsun}{L$_\odot$}
\newcommand{\Mhi}{$M_{\rm HI}$}
\newcommand{\Mst}{$M_\star$}
\newcommand{\must}{$\mu_\star$}
\newcommand{\nuvr}{NUV$-r$}
\newcommand{\Rinz}{$R_{50,z}$}
\newcommand{\Ropt}{$R_{\rm opt}$}
\newcommand{\sov}{$S_{0.5}$}
\newcommand{\vrot}{$V_{\rm rot}$}
\newcommand{\vs}{$V_{\rm rot}/\sigma$}
\newcommand{\cindx}{$R_{90}/R_{50}$}
\newcommand{\rhalf}{$R_{50}$}
\newcommand{\tmax}{$T_{\rm max}$}
\newcommand{\gi}{$g-i$}

\newcommand{\highz}{H{\sc i}GHz}

\newcommand{\bc}{}

\title[HIGHz: A Survey of the Most HI-Massive Galaxies at z \about 0.2]{\highz: A Survey of the Most
\hi-Massive Galaxies at z \about 0.2}

\author[B. Catinella \& L. Cortese]
{Barbara Catinella$^{1}$\thanks{bcatinella@swin.edu.au} and Luca Cortese$^{1}$\\
$^{1}$Centre for Astrophysics \& Supercomputing, Swinburne University of Technology, Hawthorn, VIC 3122, Australia
}

\date{}

\maketitle

\label{firstpage}

\begin{abstract}
We present the results of the \highz\ Arecibo survey, which measured the 
\hi\ content of 39 galaxies at redshift $z>0.16$ selected from 
the Sloan Digital Sky Survey. These are all actively star-forming, disk-dominated 
systems in relatively isolated environments, with stellar and \hi\ masses larger than
$10^{10}$ \Msun\ and redshifts $0.17\leq z\leq 0.25$. Our sample
includes not only the highest-redshift detections of \hi\ emission from 
individual galaxies to date, but also some of the most \hi-massive systems known.
Despite being exceptionally large, the \hi\ reservoirs of these galaxies are 
consistent with what is expected from their ultraviolet and optical properties.
This, and the fact that the galaxies lie on the baryonic Tully-Fisher relation,
suggests that \highz\ systems are rare, scaled-up versions of local
disk galaxies. 
We show that the most \hi-{\bc massive} galaxies discovered in the Arecibo Legacy Fast ALFA
survey are the local analogues of \highz, and discuss the possible connection between 
our sample and the turbulent, gas-rich disks identified at $z\sim 1$.
The \highz\ sample provides a first glimpse into the properties of the
massive, \hi -rich galaxies that will be detected at higher redshifts by the 
next-generation \hi\ surveys with the Square Kilometer Array and its pathfinders.
\end{abstract}

\begin{keywords}
galaxies: kinematics and dynamics -- galaxies: evolution -- 
galaxies: fundamental parameters -- radio lines: galaxies
\end{keywords}

\section{Introduction}\label{s_intro}

In the past decade or so, studies of the cold atomic (\hi) and 
molecular (\htwo) hydrogen content of local galaxies have progressed
from observations of relatively small samples to large surveys,
including from several hundreds to a few tens of thousands objects
\citep{hipass,alfalfa,gass1,coldgass1}. 
This significant boost in number statistics, as well as in 
data quality, has made it possible to characterize the 
cold gas properties of galaxies from a statistical point of view,
and to identify the most important scaling relations connecting gas 
and other galaxy properties in the local Universe 
\citep{gass1,coldgass1,luca11,huang12,boselli14b}.

One of the main challenges for \hi\ and \htwo\ astronomy is 
now to extend these investigations to cosmological distances, probing
the variation of the cold gas content of galaxies with the age of the Universe. 
This is particularly important, as the remarkable decrease in the cosmic star formation 
rate density from $z\sim 1$ to 0 \citep[\eg,][]{lilly96, madau96,bell05}
must be a direct consequence of a change in the gas cycle of galaxies. 

Thanks to the upgrade of the IRAM Plateau de Bure interferometer, 
observations of molecular hydrogen (as traced by the CO lines) 
have recently been able to reach redshifts
that were unimaginable a decade ago, gradually revealing 
a population of exceptionally gas-rich ($M_{H_2}$/\Mst \about 1), turbulent disk galaxies 
at $z= 1$-2  \citep{daddi10,genzel11,phibss}. Although only upcoming surveys with 
the Atacama Large Millimeter Array will determine whether or not these are 
representative of the star-forming galaxies at those redshifts, or
just the gas-rich tail of the distribution, these pioneering observations
confirm that the physical conditions of the interstellar medium 
have changed significantly in the last $\sim$8 billion years, and 
highlight the importance of the very gas-rich regime for galaxy 
evolution studies. 

Unfortunately, \hi\ astronomy lags behind in this respect.
Due to sensitivity of current instruments, as well as man-made radio interference, 
\hi\ observations are still struggling to detect the weak 21 cm emission
beyond $z\sim$0.16 \citep{verheijen07,highz_pilot,jaffe12,chiles_pilot}.
Thus, spectral stacking of optically-selected samples that are undetected in \hi\ surveys
is currently the most powerful technique to measure the average \hi\ content
of galaxies at higher redshifts. Although not a
substitute for \hi\ detections, stacking has provided estimates of the cosmic \hi\ 
density up to a redshift $z\sim 0.4$ \citep{lah09}.
Only the next generation, deep \hi\ surveys with the Square Kilometer Array \citep[SKA,][]{ska} 
and its pathfinders, ASKAP \citep{askap} and MeerKAT \citep{meerkat}, will be
able to detect \hi\ emission at these and higher redshifts.
As theoretical models predict a different evolution in the 
properties of the atomic and molecular gas phases (\eg\ \citealt{obreschkow09,lagos11}), quantifying how 
the \hi\ reservoirs of galaxies vary with redshift is of primary importance.

\begin{table*}
\caption{\label{t_runs}Arecibo observing runs}
\begin{tabular}{lllcl}
\hline\hline
Project ID & Dates         & Redshift  & SDSS selection & Allocation \\
\hline
A1803f     & 2003 Oct 8-18; Nov 8,10,22,23; Dec 6 & 0.09 -- 0.18 & DR1$^{a}$ &  62h (18h @ $z>0.16$) \\
A1803      & 2004 Mar 8-10,24-28; Apr 11-16,25    & 0.09 -- 0.25 & DR1$^{a}$,DR2$^{b}$ & 121h (110h @ $z>0.16$) \\
A2008      & 2005 Apr 19-26                       & 0.16 -- 0.25 & DR3$^{c}$ &  72h \\
A2270      & 2007 Mar 8-16                        & 0.16 -- 0.25 & DR5$^{d}$ &  80h \\
A2428      & 2011 May 3-15                        & 0.16 -- 0.32 & DR7$^{e}$ &  76h \\
\hline
\end{tabular}
\begin{center}
{\bc $^{a}$\citet{sdss1}; $^{b}$\citet{sdss2}; $^{c}$\citet{sdss3}; $^{d}$\citet{sdss5}; $^{e}$\citet{sdss7}.}
\end{center}
\end{table*}

Despite current constraints, carefully planned observations with existing radio telescopes 
can already peek into the $z=0.2$ \hi\ Universe, providing us with a glimpse of
the gas-rich galaxy population that will be detected by the SKA pathfinders,
and whose physical conditions might resemble those of higher redshift systems. 
Because of its exquisite sensitivity, the Arecibo radio telescope
has a critical advantage for the detection of weak \hi\ signals. With a
collecting area that is equivalent to one-tenth of the full SKA, Arecibo
can already detect \hi\ emission at $z\sim 0.2$ with integrations of less than ten hours
per object, as opposed to the few hundred hours necessary with current interferometers.
Naturally, the downside of single-dish observations is their lack of spatial resolution,
hence detecting \hi\ emission at these redshifts is practically restricted to
carefully selected galaxies in low density environments. However this drawback
is partly compensated by the fact that gas-rich galaxies (which are the most 
likely to be detected) are preferentially found in isolation \citep{papastergis13}. Indeed, the feasibility
of these observations was demonstrated by our pilot survey \citep{highz_pilot}.

In this paper we present the completed \highz\ survey, which measured the \hi\
content of 39 optically-selected galaxies in the $0.17 <z< 0.25$ redshift interval. 
We describe our entire observing campaign and investigate the properties of 
the detected galaxies. In addition to the the highest redshift detection 
of \hi\ emission from a galaxy to date ($z=0.25$), this sample
includes some of the most \hi\ massive galaxies currently known.
We discuss the relevance of \highz\ in the context of other surveys of
exceptionally gas-rich galaxies, both in the local and in the higher redshift Universe.

All the distance-dependent quantities in this work are computed
assuming $\Omega=0.3$, $\Lambda=0.7$ and $H_0 = 70$ \kmsm. 
AB magnitudes are used throughout the paper.

\section{Sample selection and Arecibo observations}\label{s_sample}

As mentioned in \citet{highz_pilot}, where we presented our initial results,
this programme started as a pilot survey to detect \hi\ emission 
from disk galaxies at $z>0.05$, \ie\ beyond the redshift of past Arecibo 
surveys. We soon realized that the frequency interval corresponding to 
$0.11<z<0.16$ was inaccessible because of 
Radio Frequency Interference (RFI), therefore we concentrated
our efforts on the $z>0.16$ targets. By then, the first 
data release \citep{sdss1} of the Sloan Digital Sky Survey \citep[SDSS;][]{sdss}
became available, which provided the ideal data base to 
search for galaxies with potentially large \hi\ content, although the 
spectroscopic coverage of the sky area accessible to Arecibo was very limited.

The targets for \hi\ spectroscopy were selected from the most
recent SDSS spectroscopic data release available at the time of the observations
(see Table~\ref{t_runs}) and according to the following criteria: 
(a) objects spectroscopically classified as galaxies;
(b) observable from Arecibo during night-time (to minimize the impact
of RFI and solar standing waves on our data);
(c) redshift $0.16<z<0.27$. This interval corresponds to frequencies
from 1120 to 1220 MHz and is set by a filter; 
(d) inclination $i \geq 45$\deg\ {\bc (computed from the axis ratio in \rband\ as in
\citealt{gass1})}, for use in disk scaling relations. In the most 
recent run we relaxed this condition to include inclinations $i \geq 30$\deg;
(e) presence of \Ha\ emission in the SDSS fiber, with line width between 
100 and 700 \kms\ and equivalent width between 5 and 50 Angstr{\"o}m
{\bc (to avoid extreme star formation rates, usually associated to merger
systems and/or starburst galaxies)};
(f) exponential disk profile, based on the likelihood of exponential versus
DeVaucouleurs fit to the \rband\ profile (likelihood(exp)/likelihood(DeV)$>10^6$).

After compiling the candidate list for each run, we discarded galaxies with
redshifts corresponding to the frequency of known RFI, and/or in the 
vicinity of NVSS continuum sources that would cause ripples in the baselines.
We then carefully inspected the SDSS image of each remaining galaxy, and excluded 
those with interacting or peculiar appearance, and/or with \hi\ emission possibly 
contaminated by that of nearby objects. Specifically, we discarded targets 
with galaxies of similar size or luminosity lying within a 4\arcmin\ radius 
(twice the the Arecibo beam). This was a crucial step in order to minimize the 
likelihood that our \hi\ measurements are contaminated by companions with
{\it potentially comparable HI content}, as the half power full width of the beam at the 
frequencies of our observations, \about 4\arcmin, subtends \about 800 kpc at $z=0.2$.
From the final list, we gave priority to the galaxies that looked more promising,
\ie\ those with largest apparent size and/or presence of spiral arms.
Hence, although the parent sample out of which the targets were extracted is
volume-limited and well defined, we deliberately picked our galaxies one by one,
in order to maximize our chances of detecting \hi\ emission. Not surprisingly,
we obtained a sample that is strongly biased toward \hi-{\bc massive} objects.
{\bc We targeted 49 galaxies with effective on-source integration times 
ranging between 52 and 260 minutes, and detected 39 objects. 
The 10 galaxies that were not detected have redshifts between
0.23 and 0.26 (except one with $z=0.17$). A number of other
potential targets were observed for a short time and abandoned because of various
issues (too close to RFI, continuum sources, bad baselines).}

The \hi\ data were collected under four Arecibo programs and during several 
observing runs, which took place between Fall 2003 and Spring 2011. 
Table~\ref{t_runs} lists the Arecibo program identifiers, the dates of the observations, 
the targeted redshift interval, the SDSS data release used for target selection,
and the amount of telescope time allocated. The A1803 program was split into
a Fall and a Spring portion (we indicate the former as A1803f), and was partly
devoted to lower redshift observations. The A2428 program was accepted in 2008 
but scheduled on the telescope three years later.
Including all the overheads, the total time spent on the $z>0.16$ targets was
356 hours. 

The observations were all carried out on site and
in standard {\em position-switching} mode: each
observation consisted of an {\it on/off} source pair, each integrated
for 4 minutes (5 minutes in the last two observing runs), followed by the
firing of a calibration noise diode. We used the L-band wide receiver,
which operates in the frequency range 1120$-$1730 MHz, with a
1120$-$1220 MHz filter and a 750 MHz narrow-band (60 MHz) front-end filter 
to limit the impact of RFI on our observations.
The interim correlator was used as a backend, and the
spectra were recorded every second with 9-level sampling. 
Two correlator boards, each configured for 12.5 MHz
bandwidth, one polarization, and 2048 channels per spectrum
(yielding a velocity resolution of 1.8 \kms\ at 1200 MHz before
smoothing) were centered at or near the frequency
corresponding to the SDSS redshift of the target. 
The Doppler correction for the motion of the Earth was applied during
off-line processing.   
We note that, apart from the different redshift interval, the observing
setup and data reduction pipeline are identical to those used for 
the GALEX Arecibo SDSS Survey \citep[GASS;][]{gass1}.

\section{Data reduction}

The data reduction was performed in the IDL environment using our own
routines, which are based on the standard Arecibo data processing
library developed by Phil Perillat.
In summary, the data reduction of each polarization and
{\it on-off} pair includes Hanning smoothing, 
bandpass subtraction, RFI excision, and flux calibration.
Once the data are flux calibrated, a total spectrum is obtained for each of the
two orthogonal linear polarizations by combining
good quality records (those without serious RFI or standing waves).
Each pair is weighted by a factor $1/rms^2$, where $rms$ is the root mean square
noise measured in the signal-free portion of the spectrum. 
The two polarizations are separately inspected and averaged, yielding the
final spectrum.

After boxcar smoothing and baseline subtraction, the \hi-line
profiles are ready for the measurement of redshift, rotational
velocity and integrated  \hi\ line flux. 
Recessional and rotational velocities are measured at the 50\%
peak level from linear fits to the edges of the \hi\ profile.
Our measurement technique is explained in more detail, \eg, in
\citet[][\S 2.2]{widths}.

Below we describe in more detail two very important steps of our data
reduction, RFI excision and flux calibration.

\begin{figure}
\begin{center}
\includegraphics[width=8.5cm]{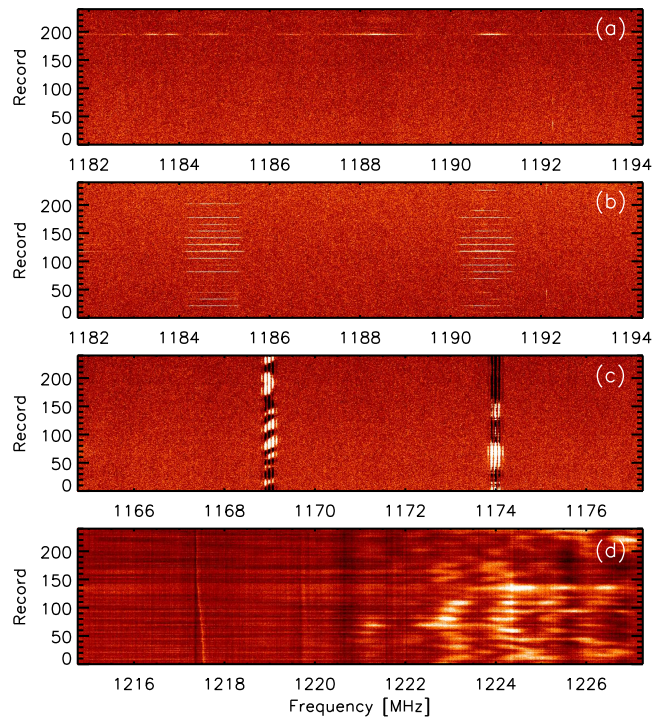}
\caption{Examples of RFI affecting our data (from the A1803 run). Each panel shows the time-frequency 
representation of a processed on-off pair for the first correlator board. Spectra are recorded every second,
thus the vertical scale can be read as time in seconds; the full 12.5 MHz bandwidth is shown on the horizontal axis. 
The top two panels illustrate examples of RFI that can be easily excised (see also Fig.~\ref{rfi_exc}). The RFI 
in (c) cannot be removed, but most of the frequency bandpass is still useful for 
the observations; (d) shows an observation that is completely compromised.}
\label{rfi}
\end{center}
\end{figure}

\begin{figure*}
\begin{center}
\includegraphics[width=17.5cm]{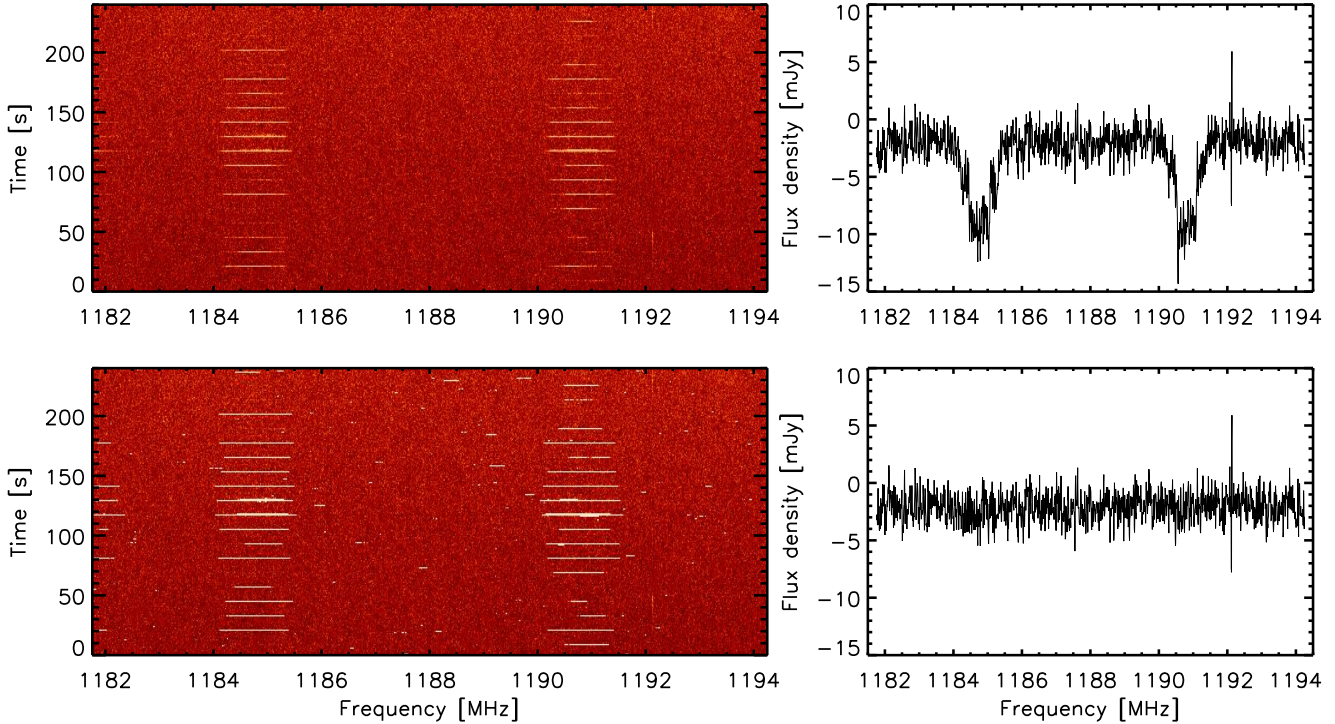}
\caption{RFI excision. {\it Top:} Time-frequency representation of an observation of AGC 212887, showing 
strong RFI near 1185 and 1191 MHz (left), and corresponding spectrum (right). The RFI is in the off-source
observation, thus it appears as absorption-like features in the spectrum.  {\it Bottom:} The RFI mask is 
shown on top of the data (white pixels) on the left; the corresponding spectrum is RFI-free (right), except
for a very narrow spike at 1192.1 MHz that is not identified by the RFI-masking algorithm.}
\label{rfi_exc}
\end{center}
\end{figure*}

\subsection{RFI excision}


The presence of RFI in our observations, which target a frequency interval that is 
well below the radioastronomy protected band (1400--1427 MHz), was the main challenge
for our survey.

RFI can be generated both internally (by electrical equipment such as
digital correlators and computers) and externally (\eg\ by broadcasting 
radio and television stations, mobile phones, airport radars, and telecommunication
satellites, but also by natural sources such as lightning). Single-dish radio telescopes are 
especially vulnerable to this problem because all incoming RFI, entering by scattering or 
reflection, enters the system coherently \citep{fridman01}.
Due to the variety of its possible sources, RFI spans a wide range of characteristics,
both in the time and frequency domains; it can also be strongly polarized or completely unpolarized.

A few examples of the most common types of RFI that affected our observations
are illustrated in Fig.~\ref{rfi}. Each panel is the result of a processed 4 minute
integration for the first correlator board (which shows one polarization only; the second board
is almost identical unless the interference is strongly polarized); each row is a single spectrum, 
recorded every second. We always refer to frequency {\it channels} along the $x$ axis
and {\it records} along the $y$ axis.
In (a) the RFI appears for a very short time (a few seconds at most, around
record 200), affecting all the bandpass; in (b) it is localized in the frequency domain
(between 1184 and 1186 MHz, and between 1190 and 1192 MHz), but is present in several
records. In both cases, the interference can be easily removed during data 
reduction by excising the affected records. These examples demonstrate the importance
of a fast dump rate for spectrum recording, in order to minimize the amount of data
that must be discarded because of RFI contamination. Panel (c) shows strong RFI signals
at 1169 MHz and 1174 MHz; in this case the RFI cannot be simply removed
by excising bad records, but the observation is not be compromised as
long as the interference does not sit on top of the galaxy emission.
Indeed, this example shows how critical an accurate knowledge of the galaxy redshifts
was for this project, in order to anticipate at which frequencies the expected \hi\ signal 
(which starts appearing only after many pairs have been co-added) would lie.
Lastly, the presence of strong RFI outside the 12.5 MHz bandpass in (d) makes the data unusable.
Our first observations immediately showed that the 1220--1230 MHz 
and 1270--1280 MHz frequency intervals were inaccessible due to type (d) RFI.

Accurate planning of our observations was essential in order to minimize the impact of RFI on our data.
As mentioned in Section~\ref{s_sample}, in addition to avoiding targets whose \hi\ emission would lie
in the proximity of known RFI of the type shown in Fig.~\ref{rfi}c, we used
front-end bandpass filters to restrict as much as possible the frequency interval
``seen'' by the correlator (to avoid RFI of the type of Fig.~\ref{rfi}d).

Ideally, it would be convenient to rely on automatic algorithms to identify and remove RFI, but
defining properties that distinguish RFI from astronomical signals is not a trivial task,
especially in the presence of weak interference. Such algorithms are generally based on thresholding 
techniques, by which a portion of the data in the time and/or frequency domain is discarded when 
its mean (or some other statistical indicator) exceeds a certain value. A thorough testing 
is usually necessary to determine how to set the thresholds.
Contrary to automatic algorithms, the human eye can easily identify RFI from the inspection of 
time-frequency representations such as those shown in Fig.~\ref{rfi} and thus, although 
impractical, manual excision achieves much better results. Because reliable RFI excision was 
extremely important for our survey, we resorted to an admittedly time-consuming, hybrid approach. 

RFI excision is applied to the data before flux calibration, and its output is an RFI mask, which records the  
pixels in the time-frequency domain marked for rejection for each polarization and pair.
The process includes the following steps:

(1) Processing each polarization of the {\it on-off} pair. Because we process the pairs first,
RFI appears as an emission or absorption feature in the spectra depending on whether 
it is present/stronger in the {\it on} or in the {\it off} scan, respectively (see Fig.~\ref{rfi_exc}).
This also implies that we discard data when RFI is present in at least one of the {\it on/off} scans.

(2) For each frequency channel, an {\it rms} noise spectrum across the time direction is computed, and an iterative
linear fit performed, rejecting any points that deviate by 3$\sigma$ or more from the fit. 
Fitting of the same channel is repeated until no more points are rejected, or an iteration limit is reached.
Optionally, the data can be smoothed over $N_{sm}$ frequency channels before searching for RFI. For our data 
set, we found that the best solution was to smooth by 0.3 MHz, and we did this by default.

(3) The rejected points are flagged, and time-frequency images for the two polarizations with
the RFI mask overlaid (along with histograms showing the noise in each record) are displayed. This is illustrated 
in the bottom left panel of Fig.~\ref{rfi_exc} for one polarization of one of our observations.

(4) At this point we inspect how well the automatic excision worked, and either run the program again
or move to the next step. It is often necessary to run the program a few times with a different sigma 
threshold for pixel rejection and/or changing the number of channels $N_{sm}$  for spectral smoothing
in order to obtain the best result. RFI of the type seen in Fig.~\ref{rfi}a usually requires to
mask entire records, as many of the pixels in the affected records deviate by less than 3$\sigma$
from the average noise. This is done by choosing a threshold $r$ for record rejection 
(records with $r$ per cent or more bad pixels are flagged).

(5) If necessary, we perform manual excision by selecting regions of the time-frequency image 
that should be rejected. The selected region is added to the RFI mask 
generated in point 4. Manual excision is usually needed for weak RFI that is not adequately 
masked by the automatic algorithm, and for unusual RFI (\eg, drifting in frequency).

(6) Lastly, the spectrum obtained after applying the RFI mask is shown (Fig.~\ref{rfi_exc}, bottom right).
Here we can go back to (1) if needed, reject the pair if the data are unusable, or proceed
to the flux calibration (the pair is reprocessed using the final RFI mask), keeping track 
of the number of records used at each frequency.

This algorithm is designed to identify strong RFI features in the time domain
and spectrally broad RFI which is time-variable. RFI that is present in all records
at a given frequency, such as that in Fig.~\ref{rfi}c, is not masked. 
Fig.~\ref{rfi_exc} shows an example of the application of our
automatic RFI excision algorithm to our data, and the resulting \hi\ spectrum before
and after RFI masking (right panels).

\subsection{Flux calibration}\label{s_calib}

Standard position-switched observations of a source with Arecibo are followed
by two measurements of a calibration diode, which is turned on for 10 seconds and then off 
for another 10 seconds, while the telescope points to the blank sky 
(the {\it off} galaxy position). The noise diode has a known temperature as
a function of frequency for each polarization, hence processing its {\it on/off} observations
allows one to determine the ``system temperature'', and convert correlator units into 
Kelvin degrees. The conversion to spectral flux density units is achieved by
applying the gain curve, which provides the point source gain of the telescope in
K/Jy for the specific receiver as a function of frequency and zenith angle of 
the observation. As discussed in \citet{vanzee97,springob05} and most recently in 
\citet[][see their section 5.2]{alfalfa40}, \hi\ line flux densities derived
from targeted {\bc single-dish} observations in modern data sets are typically accurate to not
better than 15\%.

A technical problem made all the calibration scans taken between March 26 2004 and
May 2005 unusable, thus we had to resort to a non-standard flux calibration for
part of the A1803 and all the A2008 data. Because of a bug in the telescope control software 
(fixed in June 2005), selecting the 750 MHz narrow-band filter also silently triggered the 
``winking'' calibration mode, which is used for pulsar observations. This caused
our noise diode to be switched on and off with a frequency of 25 Hz (\ie, every 40 ms),
which had two effects: (a) render our calibration scans useless, and (b) inject extra
noise in our observations, as the diode was on half of the time. This effectively
increased the system temperature $T_{sys}$ by \about 15\% (estimated from
$0.5 \times T_{diode}/T_{sys} \sim 1.15$, where $T_{diode}$ and
$T_{sys}$ are approximately 10 K and 33 K for our observations, respectively).

In order to calibrate these data, Phil Perillat at Arecibo kindly provided us with System 
Equivalent Flux Density (SEFD) curves for our receiver, which are obtained from a fit to the
system temperature as a function of gain of the telescope. These fits give the point source 
SEFD of the telescope in Jy, as a function of frequency and zenith angle of the observation.
We processed all the good calibration scans taken in 2003 and 2004, compared the
diode calibration with the SEFD one, and found the two to be consistent within less than 10\%.
The SEFD curves are based on archival measurements of the system temperature taken under
standard conditions, \ie\ $T_{sys} \sim 33$~K at the frequencies of \highz. Hence, in order
to account for the increased system temperature caused by the winking calibration,
we multiplied the conversion factor obtained from the SEFD curves by 1.15.
Conservatively, we consider the flux calibration of the galaxies affected by the winking
calibration to have an additional uncertainty of 15\% over the standard calibration, or 
a total uncertainty of \about 21\%. This problem affected partly or completely 20 out of 39 galaxies,
however the larger uncertainty in those \hi\ fluxes does not affect our conclusions at all.

\begin{figure}
\begin{center}
\includegraphics[width=8.2cm]{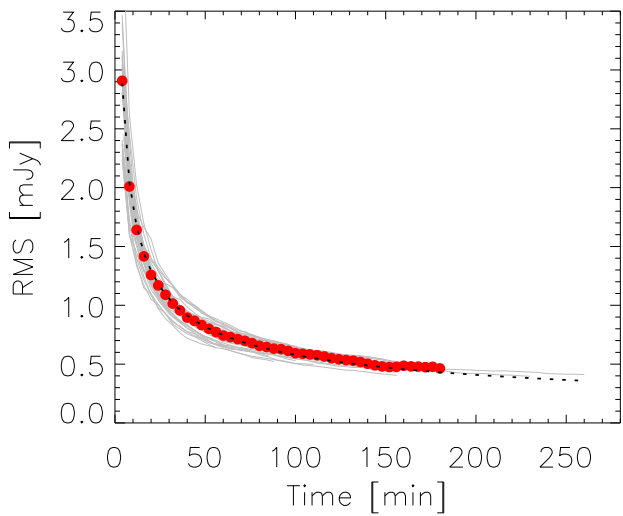}
\caption{The {\it rms} noise is plotted versus on-source integration time, $T_{int}$, for individual galaxies 
(gray lines); red circles are running averages (computed when at least three data points were available). 
The {\it rms} decreases as $\sqrt{T_{int}}$ as expected (dotted line).}
\label{rms}
\end{center}
\end{figure}

To our knowledge, these are the longest observations of individual galaxies done with
Arecibo. As can be seen in Fig.~\ref{rms}, the measured {\it rms} noise of our observations computed
from the unsmoothed \hi\ spectra (\ie\ at a velocity resolution of 1.8 \kms\ at 1200 MHz) decreases as 
the square root of the integration time, as expected.

\section{Data presentation}

\subsection{\hi\ catalogue}

The measured \hi\ parameters for the 39 detected
galaxies are listed in Table~\ref{t_det}, ordered by increasing right ascension:\\
Col. (1): identification code in the Arecibo General Catalog (AGC,
maintained by M.P. Haynes and R. Giovanelli at Cornell University).\\
Col. (2): SDSS identifier. \\
Col. (3): on-source integration time of the Arecibo observation, $T_{\rm on}$, in minutes. This 
number refers to {\it on scans} that were actually combined, and does not account for
losses due to RFI excision {\bc (typically of order of a couple of percent). The fraction of 
usable data for our sample, \ie\ the ratio between $T_{\rm on}$ and the total on-source time 
actually spent on each galaxy, varied between 50\% and 100\%, with an average of 84\%. 
Pairs were discarded because of RFI that could not be excised or bad baselines.}\\
Col. (4): velocity resolution of the final, smoothed spectrum in \kms. \\
Col. (5): redshift, $z$, measured from the \hi\ spectrum.
The error on the corresponding heliocentric velocity, $cz$, 
is half the error on the width, tabulated in the following column.\\
Col. (6): observed velocity width of the source line profile
in \kms, \whi, measured at the 50\% level of each peak. 
The error on the width is the sum in quadrature of the 
statistical and systematic uncertainties in \kms\ (the latter
depend on the subjective choice of the \hi\ signal boundaries
and are usually negligible, see \citealt{gass1} for details).\\
Col. (7): velocity width corrected for instrumental broadening
and cosmological redshift only, \whi$^c$, in \kms\ (see \citealt{gass_dr2}). 
No inclination or turbulent motion corrections are applied.\\
Col. (8): observed, integrated \hi-line flux density in Jy \kms,
$F \equiv \int S~dv$, measured on the smoothed and baseline-subtracted
spectrum. The reported uncertainty is the sum in quadrature of the 
statistical and systematic errors (see \citealt{gass1} for details).\\
Col. (9): {\it rms} noise of the observation in mJy, measured on the
signal- and RFI-free portion of the smoothed spectrum.\\
Col. (10): signal-to-noise ratio of the \hi\ spectrum, S/N,
estimated following ALFALFA and GASS (see \eg\ \citealt{gass1}).\\
Col. (11): base-10 logarithm of the \hi\ mass, \Mhi, in solar
units, computed via: 
\begin{equation}
    \frac{M_{\rm HI}}{\rm M_{\odot}} = \frac{2.356\times 10^5}{1+z}
    \left[ \frac{d_{\rm L}(z)}{\rm Mpc}\right]^2
    \left(\frac{\int S~dv}{\rm Jy~km~s^{-1}} \right)
\label{eq_MHI}
\end{equation}
\noindent
where $d_{\rm L}(z)$ is the luminosity distance to the galaxy at
redshift $z$ as measured from the \hi\ spectrum. \\
Col. (12): base-10 logarithm of the \hi\ mass fraction, \Mhi/\Mst.\\
Col. (13): quality flag, Q. Code 1 indicates detections with a S/N ratio 
of 6.0 or higher. Code 2 is assigned to lower S/N, but still secure detections, 
and code 3 to marginal detections. The separation between codes 2 and 3
is not simply by S/N, but takes into account \hi\ profile and baseline
quality --- code 3 detections have more uncertain \hi\ parameters, hence 
are shown with different symbols in our plots.
In all cases, the \hi\ redshift is consistent with the SDSS measurement.\\
Col. (14): Arecibo project identifier (see Table~\ref{t_runs}).\\

{\bc  For the 10 galaxies that were not detected (see section 2) we 
did not obtain stringent constraints on the \hi\ masses --- the upper limits
are comparable to or higher than the \hi\ masses of the detections. This is
because, after the initial 1 or 2 hours of on-source integration, we completed
the observations of the target only if there was a hint of galaxy \hi\ emission 
at the expected frequency.}\\

SDSS postage stamp images and \hi\ spectra of the \highz\ {\bc detections}
can be found in the Appendix. Fig.~\ref{hist} presents the distributions
of measured redshifts, rotational velocities and \hi\ masses for both
high-quality (hatched) and marginal (solid histograms) detections.
{\bc The \highz\ galaxies have \hi\ masses that vary between 1.9 and 7.9 $\times 10^{10}$ \Msun,
at the top end of the \hi\ mass function (\hi MF; \citealt{himf_aa}).
Because the \hi MF drops rapidly above $\log M_{\rm HI,*}/$M$_\odot = 9.96$,
these systems are rare in the local Universe (for instance, galaxies with
\Mhi $\geq 5 \times 10^{10}$ \Msun\ are over a factor of 100 less
common than $M_{\rm HI,*}$ ones).}

\begin{figure*}
\begin{center}
\includegraphics[width=17.5cm]{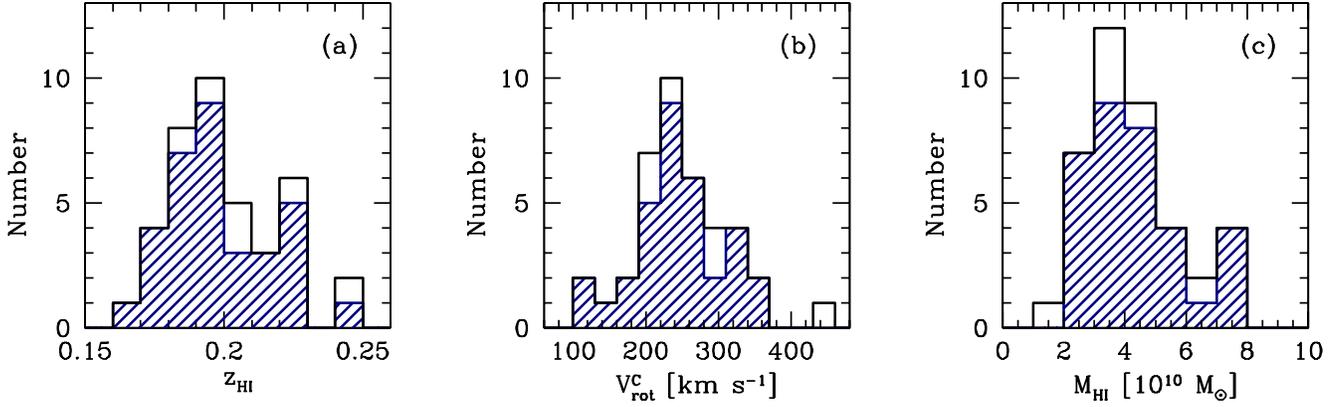}
\caption{Distributions of (a) redshifts, (b) rotational velocities, and
(c) \hi\ masses for this sample (solid histograms). Hatched histograms
do not include the marginal detections.}
\label{hist}
\end{center}
\end{figure*}

\subsection{SDSS and GALEX data}\label{s_sdss}

This section summarizes the quantities derived from optical and UV
data used in this paper. All the optical parameters listed below were
obtained from Structured Query Language (SQL) queries to the SDSS
DR7 database server\footnote{
{\em http://cas.sdss.org/dr7/en/tools/search/sql.asp}
},
unless otherwise noted.

The NUV magnitudes for our sample were obtained from 
the {\it GALEX Unique Source Catalogs}\footnote{
{\em http://archive.stsci.edu/prepds/gcat/}
} \citep{seibert12}.
The measured \nuvr\ colors are corrected for Galactic
extinction following \citet{wyder07}, from which we obtained
$A_{NUV}-A_r = 1.9807 A_r$ (where the extinction $A_r$ is available
from the SDSS data base and reported in Table~\ref{t_sdss} below). We
did not apply internal dust attenuation corrections.

Table~\ref{t_sdss} lists the relevant SDSS and UV quantities for the galaxies
published in this work, ordered by increasing right ascension:\\
Cols. (1) and (2): AGC and SDSS identifiers. \\
Col. (3): SDSS redshift, $z_{\rm SDSS}$. The average uncertainty of
SDSS redshifts for this sample is 0.00015.\\
Col. (4): base-10 logarithm of the stellar mass, \Mst, in solar
units. Stellar masses are from the MPA/JHU SDSS DR7 catalog\footnote{
{\em http://www.mpa-garching.mpg.de/SDSS/DR7/}; we used the improved stellar masses from
{\em http://home.strw.leidenuniv.nl/~jarle/SDSS/}
}, and are derived from SDSS photometry using the
methodology described in \citet[][a \citealt{chabrier03}
initial mass function is assumed]{salim07}. In one case, we
replaced the SDSS stellar mass estimate with a different one (AGC~232041, see below).\\
Cols. (5) and (6): radii containing 50\% and 90\% of the Petrosian
flux in \rband, $R_{50}$ and $R_{90}$ respectively, in arcsec.\\
Col. (7): radius containing 90\% of the Petrosian flux in \rband, $R_{90}$,
in kpc.\\
Col. (8): base-10 logarithm of the stellar mass surface density, \must, in
\Msun~kpc$^{-2}$. This quantity is defined as 
$\mu_\star = M_\star/(2 \pi R_{50}^2)$, with $R_{50}$ in kpc units.\\
Col. (9): Galactic extinction in \rband, ext$_r$, in magnitudes, from SDSS.\\
Col. (10): \rband\ model magnitude from SDSS, $r$, corrected for Galactic extinction.\\
Col. (11): minor-to-major axial ratio from the exponential
fit in \rband, $(b/a)_r$, from SDSS.\\
Col. (12): inclination to the line-of-sight in degrees, computed from
$(b/a)_r$ in the previous column assuming an intrinsic axial ratio $q_0=0.20$
as in \citet{gass4}. \\
Col. (13): total star formation rate (SFR) in \Msun~yr$^{-1}$, from 
the MPA/JHU SDSS DR7 catalog. These SFRs are based on the technique discussed in
\citet{jarle04}.\\
Col. (14): \nuvr\ observed color, corrected for Galactic extinction.\\
Col. (15): source of the UV photometry: GALEX All-sky Imaging Survey (AIS) or 
Medium Imaging Survey (MIS; see \citealt{galex}).

All our targets have SDSS MPA/JHU stellar masses in the interval
log \Mst/\Msun $= [10.3, 11.4]$ except for AGC~232041, which has an
unreasonably large value of log \Mst/\Msun $=12.37$. Hence we computed
new stellar mass estimates from the \iband\ luminosity and the \gi\ colour
following \citet{zibetti09}, and applying k corrections based on analytical 
approximations by \citet{chilingarian10}. For the other galaxies in our sample
there is an excellent correlation between these estimates and the MPA/JHU values,
with an offset of 0.21 dex (Zibetti's values are systematically larger).
Hence we replaced the stellar mass of AGC~232041 with the new estimate,
log \Mst/\Msun $=11.05$ (after subtracting the systematic offset), which 
falls well within the stellar mass interval of the other galaxies in our sample.
This galaxy is identified by a different symbol in our plots.

\begin{figure*}
\begin{center}
\includegraphics[width=17.5cm]{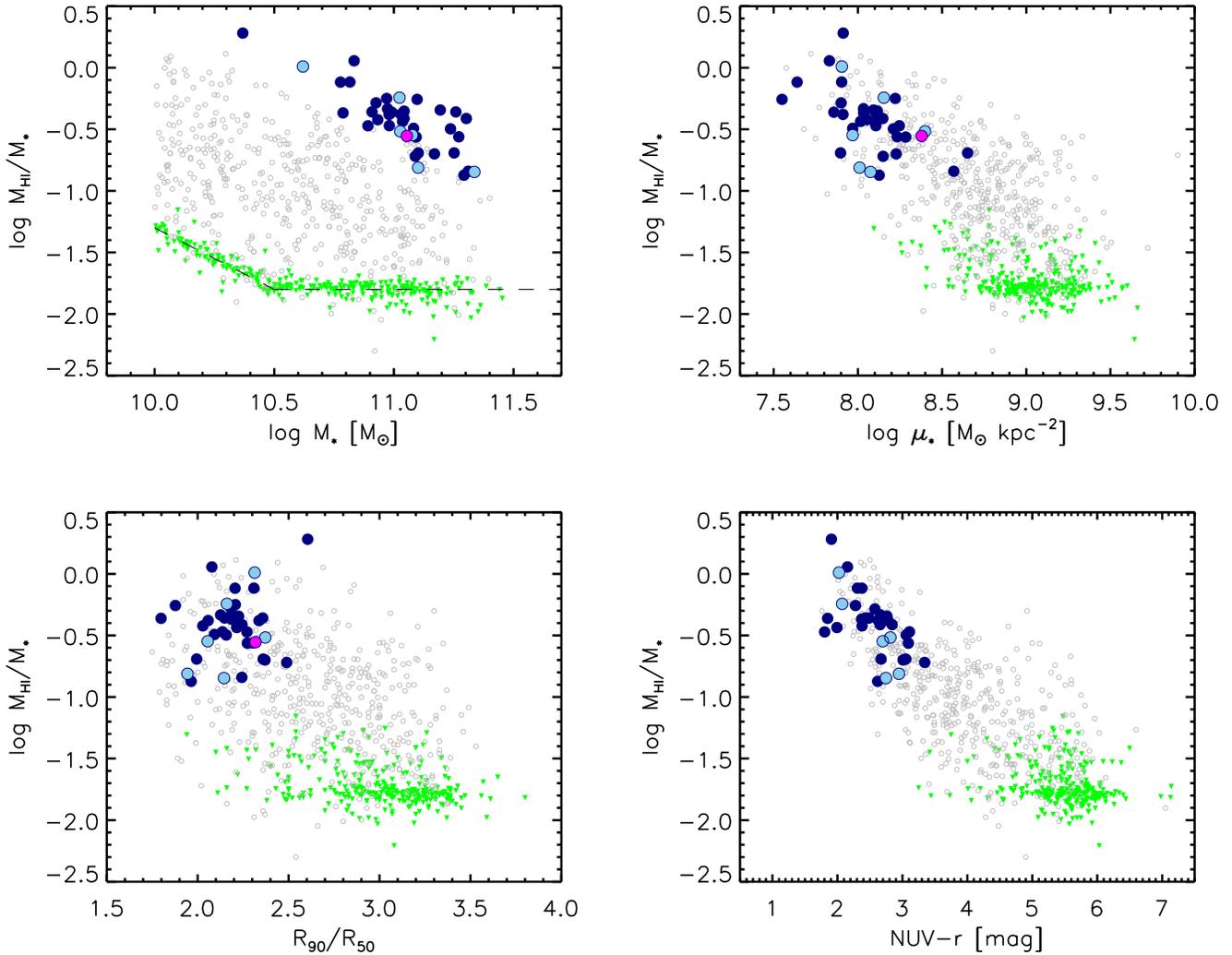} 
\caption{Gas fraction scaling relations. The \hi\ mass fraction of the sample is plotted here as
a function of stellar mass, stellar mass surface density, concentration index, and observed \nuvr\ colour
(large blue symbols; light blue circles are marginal detections, and the magenta circle indicates 
AGC~232041 -- see text). For comparison, we also show the \hi\ detections (gray dots) and the non-detections 
plotted at their upper limits (green upside-down triangles) from the GASS sample.
The dashed line in the top-left panel indicates the \hi\ detection limit of GASS.}
\label{dr3gf}
\end{center}
\end{figure*}

\begin{figure*}
\begin{center}
\includegraphics[width=17.5cm]{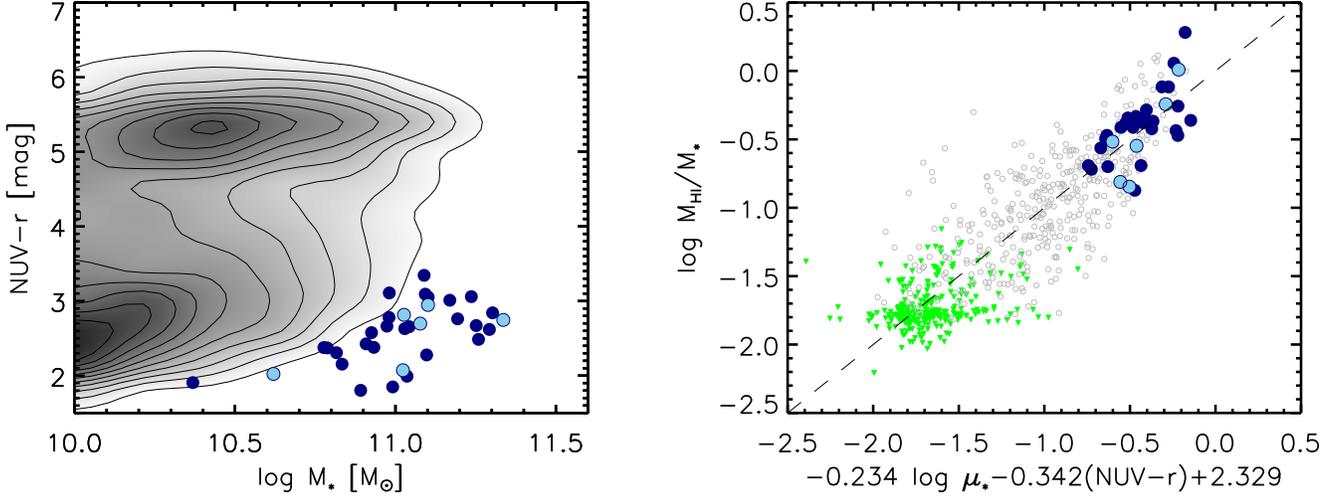}
\caption{{\it Left:} \nuvr\ colour-stellar mass diagram. Contours and grayscales 
show the distribution of the GASS parent sample for comparison; blue symbols indicate \highz\ galaxies
as in Fig.~\ref{dr3gf}. Galaxies in the \highz\ sample are unusually blue
for their stellar masses. {\it Right:} GASS gas fraction plane, showing the relation 
between measured ($y$ axis) and predicted ($x$ axis) \hi\ mass fractions; the dashed line
indicates the 1:1 relation, and the symbols are as in Fig.~\ref{dr3gf}.
The \highz\ galaxies are {\it not outliers} in this relation.}
\label{cmd}
\end{center}
\end{figure*}

\section{Results}

By selection, the \highz\ sample includes unusually \hi-rich galaxies (see Fig.~\ref{hist}).
Are these the tip of the iceberg of the $z\sim 0.2$ disk galaxy population, or just {\bc peculiar objects}? 
Are their huge gas reservoirs due to recent accretion, or are these systems 
simply scaled-up versions of local gas-rich disks?

The properties of the \highz\ systems are best understood when compared 
with those of a representative sample of local galaxies with similar stellar 
masses, for which homogeneous \hi\ measurements are available. 
Such a reference sample is uniquely provided by GASS, which measured
the \hi\ content of \about 800 galaxies selected only by 
stellar mass ($10 < \log (M_\star/M_\odot) < 11.5$) and redshift ($0.025 < z < 0.05$).
Not only GASS and \highz\ span the same stellar mass interval, but the
\hi\ observations were also taken and processed in the same way.
We use here the final data release of GASS \citep{gass_dr3} and, for comparisons that do 
not involve the gas content, the full {\it parent sample}, which is the super-set
of 12,006 galaxies meeting GASS selection criteria out of which the targets
for Arecibo observations were extracted.

We plot the \highz\ galaxies on the main gas fraction scaling relations
identified by GASS in Fig.~\ref{dr3gf}. Clockwise from the top left, we show how the 
gas mass fraction \Mhi/\Mst\ depends on stellar mass, stellar mass surface
density, observed \nuvr\ color (a proxy for specific star formation rate, or star formation 
rate per unit of stellar mass) and concentration index \cindx\ (a
proxy for bulge-to-total ratio) for both samples. \highz\ \hi\ detections 
are shown as dark blue circles, while marginal detections are presented in light blue;
the magenta circle is AGC~232041, for which we computed our own estimate of stellar mass (see
\S~\ref{s_sdss}). GASS detections and non-detections are indicated by gray dots and green 
triangles, respectively. For consistency
with \highz, we recomputed stellar mass surface densities for GASS using \rband\ Petrosian 
radii (as opposed to \zband, as in the GASS data release papers).

The gas fraction versus stellar mass plot confirms that the \highz\ sample is 
extremely \hi-rich: these galaxies have \Mhi/\Mst\ ratios that are a factor 
$\sim$10 higher than the average value of GASS at the same stellar masses. 
{\bc These large gas fractions are comparable to those of Malin~1
(log \Mhi/\Msun $= 10.82$, log \Mst/\Msun $= 10.88$, \citealt{lelli10,HIghMass}) 
and  HIZOA~J0836-43 (log \Mhi/\Msun $= 10.88$, log \Mst/\Msun $= 10.64$, \citealt{cluver10}), 
two of the most \hi-massive galaxies known.}

Interestingly, however, the large difference between {\bc \highz\ and GASS samples} completely 
disappears if we consider the other scaling relations shown in Fig.~\ref{dr3gf}.
At fixed stellar mass surface density and \nuvr\ colour, \highz\ galaxies lie 
exactly on the trends followed by local massive galaxies. 
This automatically implies that, at fixed stellar mass, \highz\ galaxies should 
be outliers in the colour stellar mass diagram (and in the \must-\Mst\ plot, see
next section). Indeed, as shown in the left panel of Fig.~\ref{cmd}, 
\highz\ galaxies lie outside the region that includes most of the GASS parent sample, 
indicated by the gray contours. Specifically, the $z\sim 0.2$ galaxies are unusually 
blue for their stellar masses, and similar systems are rare in the local Universe. 
As \nuvr\ is an excellent proxy for specific star formation rate, 
this implies that \highz\ galaxies formed the bulk of their stars at later times 
than the typical massive galaxy at $z=0$ (see also the next section).
We note that the peculiarity of our galaxies would not appear in an optical colour 
magnitude diagram, as all our targets lie on the optical red sequence. 
This is not surprising because, at such high stellar masses, optical colours 
do not properly trace current star formation activity \citep[\eg,][]{luca12_redsp}.

The results presented in Fig.~\ref{dr3gf} support the idea that the \hi\ content of galaxies 
is mainly driven by their colour and stellar mass surface density, with the trend 
with stellar mass being just a secondary, not physically driven relation \citep[\eg,][]{gass1,fabello1}.
This is reinforced in the right panel of Fig.~\ref{cmd}, which shows the gas fraction plane defined
by GASS. This is a relation between the gas fraction measured by our \hi\ observations
and that predicted from the \nuvr\ colors and stellar surface densities of the galaxies.
We use here the gas fraction prediction calibrated on GASS galaxies with \nuvr $\leq 4.5$ mag
(see \citealt{gass_dr3})\footnote{
The calibration of the gas fraction plane was obtained with \zband\ stellar surface densities,
hence the coefficients shown in Fig.~\ref{cmd} are strictly for \zband\ \must, but we 
plotted \rband\ \must\ instead. Because the difference between \zband\ and \rband\ calibrations 
is barely noticeable, we prefer not to provide a new version of the gas fraction plane in this
paper, and simply note that this small difference does not affect our conclusions at all.
},
which is clearly the most appropriate for \highz. Our $z\sim 0.2$ galaxies follow the same
relation identified by GASS, implying that their gas content is exactly what is expected
based on their colours and stellar mass surface densities. 

\begin{figure}
\includegraphics[width=8.2cm]{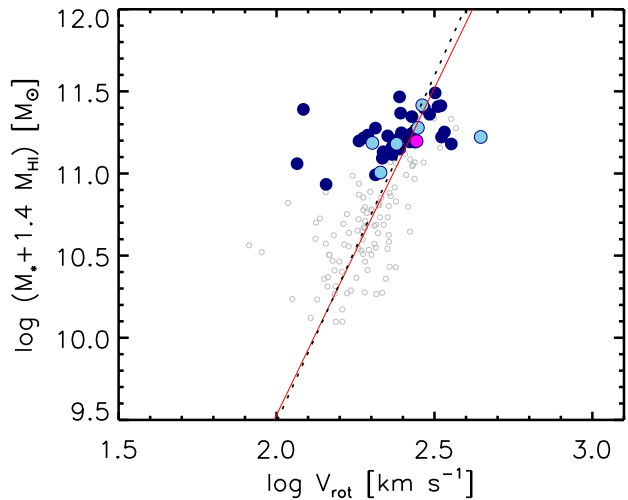}
\caption{Baryonic Tully-Fisher relation. Gray dots are the ``Tully-Fisher subset'' of inclined, 
disk-dominated GASS galaxies, reproduced from Fig.~2 of \citet{gass4}; the dotted line indicates 
the inverse fit to the GASS data points, which is in excellent agreement
with the relation from \citet[][red solid line]{mcgaugh00}. The \highz\ sample
is shown with the same symbols used in Fig.~\ref{dr3gf}.}
\label{btf}
\end{figure}

Lastly, also the rotational velocities of \highz\ galaxies are consistent with those
expected from nearby systems. This is shown in Fig.~\ref{btf}, which reproduces the baryonic 
Tully-Fisher (TF) relation for the subset of GASS galaxies with inclinations larger than 40\deg\ and \cindx $\leq 2.8$
from \citet{gass4}. Our sample meets these two criteria by selection (see Fig.~\ref{dr3gf} and 
Table~\ref{t_sdss}), except for two galaxies with inclinations slightly below
the cut (37\deg\ and 39\deg) that we show anyway. Overall, the $z\sim 0.2$ sample
lies on the baryonic TF relation, except for three outliers on the low velocity side
(from top to bottom, these are AGC 181593, 191728 and 122040) and a marginal \hi\ detection 
(AGC 249560) on the high velocity side. Notice however that there are similar outliers
at lower baryonic mass also in GASS.

To summarize, the comparison with GASS shows that
the \highz\ galaxies have unusually high gas content (in terms of both \hi\ mass and 
\hi\ gas fraction) and blue \nuvr\ colors for their stellar mass, 
but are not outliers in the other scaling relations, including the gas fraction plane. 
Their low values of stellar mass surface density and concentration index are typical of 
disk-dominated systems, and star formation appears to proceed as expected, 
despite their huge \hi\ reservoirs, with specific SFRs that are
on average one order of magnitude larger than those of GASS \hi\ detections.
In other words, all observational evidence suggests that the \highz\ galaxies 
are rare, scaled-up versions of disk galaxies in the local Universe, with 
no clear signs of peculiarity or recent interaction. 
Hence, although these objects are not spatially resolved in \hi, it seems 
reasonable to assume that their \hi\ is distributed in a disk as well, rotating at the 
expected velocity given the baryonic mass of the system.

Incidentally, the results presented above should also dispel any doubts that our \hi\ observations
might be significantly contaminated by beam confusion. This is because, if we had detected
the integrated \hi\ emission of multiple galaxies within the beam instead, the \highz\
systems would be outliers in the gas fraction plane and/or the baryonic TF relation.
This suggests that the effect of beam confusion, if present, is well within our 
observational uncertainties, as expected given our careful target selection
(see Section~\ref{s_sample}).


\section{Discussion}


Having established how \highz\ galaxies compare to the typical massive 
systems in the local Universe, we turn our attention to how these 
might fit into our general picture of galaxy evolution. 
On one side, we ask whether \hi\ reservoirs 
as large as the ones we observed are still present at $z\sim$0, despite being 
rare, or have completely disappeared. On the other side, it is natural to wonder 
if \highz\ galaxies might be the progeny of the very gas-rich turbulent disks 
observed at $z\sim$1 \citep{phibss}. 
To answer these questions, we compare \highz\ to some of the most gas-rich 
samples {\bc (with stellar masses above $10^{10}$ \Msun)} currently known between $z\sim$0 and $\sim$1.

\subsection{Are there local analogs of the \highz\ galaxies?}

As the most extreme example of massive, \hi-rich galaxies at $z=0$, 
we use the HIghMass \citep{HIghMass} sample, a set of
34 exceptionally \hi-rich local galaxies identified by ALFALFA. HIghMass galaxies
were selected from the 40\% ALFALFA data release \citep{alfalfa40} to have both
large \hi\ mass (above $10^{10}$ \Msun) and large \hi\ gas fraction for their stellar mass
(i.e. \Mhi/\Mst\ more than 1 $\sigma$ above the average, see their Fig.~1).
Other samples of massive local galaxies claimed to be ``unusually \hi-rich'' 
(\ie\ Bluedisks, \citealt{bluedisks}, and the \hi\ ``Monsters'', \citealt{HImonsters})
are, in fact, significantly less extreme than HIghMass in terms of gas fractions, 
and therefore are not shown in our plots.
For consistency with our sample, we extracted stellar masses for HIghMass from the MPA/JHU 
SDSS DR7 catalog, and kept only galaxies with $\log (M_\star/M_\odot) \ge 10$.
{\bc Stellar mass surface densities of HIghMass galaxies were computed using \rband\ 
petrosian radii and axis ratios obtained via elliptical aperture photometry of SDSS images 
(kindly provided by S. Huang, see Section 2 in \citealt{HIghMass}).}

\begin{figure*}
\begin{center}
\includegraphics[width=17.5cm]{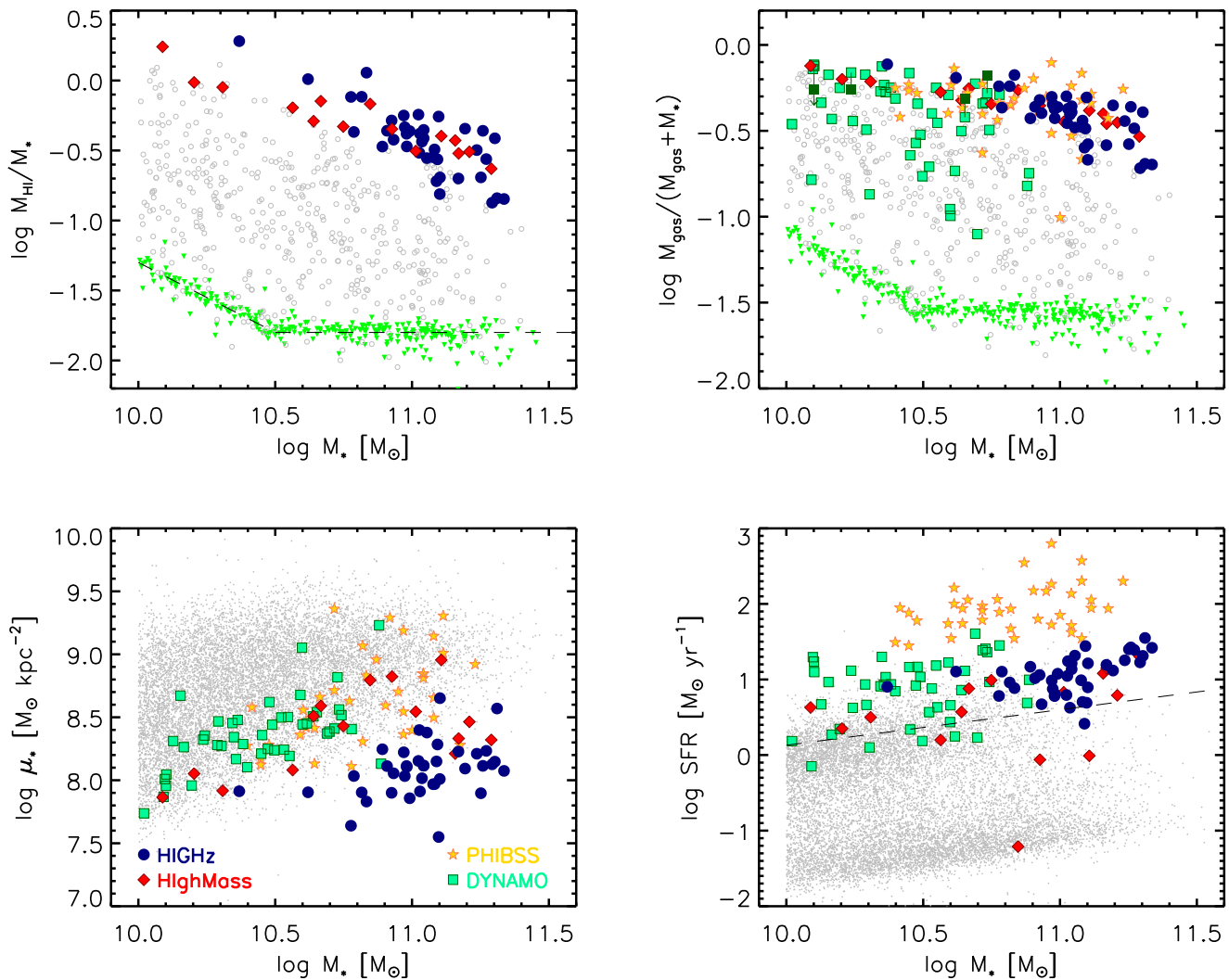}
\caption{Comparison with literature samples.
{\it Top left:} \hi\ gas fraction versus stellar mass, as in Fig.~\ref{dr3gf};
\highz\ and HIghMass galaxies with stellar masses larger
than $10^{10}$ \Msun\ are shown as blue circles and red diamonds,
respectively. {\it Top right:} Total gas fraction versus stellar mass (see text for details 
on how the total gas fractions are computed). Stars and green squares indicate PHIBSS and 
DYNAMO galaxies, respectively; dark green squares connected by lines are the four DYNAMO 
systems with available CO measurements (the leftmost one is an upper limit, as shown by the downward 
arrow). {\it Bottom left:} The stellar surface density is plotted as a function of stellar 
mass for the same data sets; gray dots show the full GASS parent sample.  {\it Bottom right:}
Star formation rate as a function of stellar mass, same symbols as bottom left panel.
A dashed line indicates the star-forming main sequence from \citet[][eq.~12]{salim07}.}
\label{hz_lit}
\end{center}
\end{figure*}

In Fig.\ref{hz_lit}, we plot both \highz\ (blue) and HIghMass (red) on some of the main scaling relations 
followed by local galaxies. It is clear that HIghMass and \highz\ galaxies have similarly 
high \hi\ gas fractions for their stellar masses (top-left panel), implying that 
$z=0$ analogs of our galaxies do indeed exist. The fact that some of 
the $z\sim 0.2$ galaxies have more extreme gas fractions is not unexpected,
as these were selected from a much larger volume (hence they are rarer)
than the ALFALFA galaxies (which have $z<0.06$). 
Moreover, both samples lie preferentially on the upper envelope of the main sequence 
of star-forming galaxies (dashed line in the bottom-right panel, from eq.~12 of \citealt{salim07}, 
which applies to galaxies with \nuvr $<4$ mag), 
confirming that their large \hi\ reservoir is 
actively feeding the formation of new stars at a significant rate, as already hinted by Fig.~\ref{cmd}.

Intriguingly, the only clear difference between \highz\ and HIghMass is in their stellar distribution (bottom-left panel):
\highz\ systems have stellar surface densities that are \about 0.6 dex smaller than HIghMass, 
implying that their optical radii are significantly larger than what observed in local, massive disks\footnote{
We checked the reliability of our \must\ measurements by taking into account
the effect of seeing on the stellar radii (the median seeing at \rband\ is 1.43\arcsec, 
according to the SDSS DR7 web site), and the conclusion remains the same, as most points in 
the plot would move upward by less than the symbol size.
}. 
This is purely a selection effect since, as mentioned in Sec.~2, we intentionally picked the galaxies 
{\it with the largest apparent sizes from the SDSS images} as targets for Arecibo observations. 
Therefore, we ended up with a sample of very uncommon galaxies, even rarer than HIghMass.

As discussed by \citealt{HIghMass}, the low \must\ values of HIghMass 
compared to the rest of ALFALFA  points out to their large halo spin parameters. 
It is a well-known prediction of galaxy formation models that stellar disks 
formed in dark matter halos with higher angular momentum content are more 
extended and {\bc have higher gas fractions} \citep[\eg,][]{mmw98,boissier00}. Hence, \highz\ galaxies are 
most likely the tail of the high spin parameter distribution identified by HIghMass.
Indeed, the gas fractions of \highz\ galaxies are consistent with 
the values predicted for spin parameters in the range $\sim$0.07-0.09 \citep{boissier00}.
This confirms that the \hi\ content, SFR and stellar distribution of our galaxies are 
what expected from their mass and angular momentum, and that no recent 
accretion and/or interaction (sometimes invoked to explain unusually gas-rich galaxies, 
\eg\ \citealt{cluver10}) are required. 

Lastly, \highz\ (and HIghMass, as discussed in \citealt{HIghMass}) systems are not 
analogs of giant low surface brightness galaxies like Malin 1 \citep{bothun87}.
Indeed, the low SFR of Malin 1 (\about 0.1 \Msun\ yr$^{-1}$, \citealt{impey89},
but see also \citealt{lelli10}) indicates that its huge \hi\ reservoir (log \Mhi/\Msun $=$10.83, 
\citealt{pickering97,lelli10}) is largely inert, whereas \highz\ galaxies are actively
star-forming, with an average SFR of 13.5 \Msun\ yr$^{-1}$ (see Table~1).
This is demonstrated by the fact that the average depletion time (\Mhi/SFR) of \highz\ systems
is $\sim$3 Gyr, in line with the typical value observed in normal star-forming galaxies 
\citep{gass2,boselli14b}.

\subsection{How do \highz\ galaxies compare with turbulent gas-rich disks at $z\sim$1?} 

As a representative sample of the {\bc high gas fraction} population of $z\sim$1 disks we 
consider PHIBSS, the IRAM Plateau de Bure HIgh-z Blue Sequence CO(3-2) Survey
\citep{phibss}. PHIBSS measured the molecular gas content of 52 star-forming galaxies 
in two redshift slices, at $z\sim 1.2$ and 2.2; most of these systems are rotationally 
supported turbulent disks. Here we restrict the PHIBSS sample to the 38 galaxies in 
the lower redshift interval ($1.00<z<1.53$), which all have CO detections.

We also include in this comparison DYNAMO (DYnamics of Newly-Assembled Massive Objects, 
\citealt{dynamo}), a survey of local star-forming galaxies selected by \Ha\ emission from the
SDSS to be potential analogs of the PHIBSS systems. 
Although for DYNAMO there are no direct \hi\ observations, and molecular gas measurements 
are available for only four objects \citep{fisher14}, given their possible connection to 
high-redshift galaxies it is very interesting to see how they relate to our \highz\ sample. 
As for HIghMass, we extract stellar masses and stellar surface densities
for DYNAMO from SDSS, consistently with our sample.

In order to compare the gas content of all samples, which have \hi\ (\highz, HIghMass) or \htwo\ (PHIBSS) 
measurements, or total gas masses simply inferred from SFRs (DYNAMO), 
we estimate total gas masses as follows.
For GASS, \highz, and HIghMass we assume a molecular-to-atomic hydrogen mass ratio 
$M_{H_2}$/\Mhi $=$0.3, which is the average value measured for a representative subset 
of GASS galaxies by COLD GASS \citep{coldgass1}. Hence, $M_{gas}= 1.768 $~\Mhi\ for these samples, 
taking into account the Helium contribution (a factor 1.36). It should be kept in mind that these are 
rough estimates, as the relation between $M_{H_2}$
and \Mhi\ in this stellar mass regime has a large scatter (0.41 dex, see Fig.~8 in \citealt{coldgass1}).
For DYNAMO, we use their published total gas content estimates, which were
inferred from the star formation rate densities using the Kennicutt-Schmidt \citep[KS;][]{kennicutt98} law,
and correct them for the Helium contribution as above.
We assume the same $M_{H_2}$/\Mhi\ ratio obtained from GASS to compute $M_{gas}$ for the four galaxies 
with available CO data. For PHIBSS, we assume that the hydrogen is all in molecular phase,
\ie\ $M_{gas}=M_{H_2}$ (\citealt{phibss}; their measurements already include the He contribution). 

The results of our comparisons are again presented in Fig.~\ref{hz_lit}. 
Keeping in mind all the assumptions in our calculations, the 
top-right panel shows that the total gas content of the bulk of $z\sim 1.2$ galaxies is comparable to 
that of both \highz\ and HIghMass, although the partition between atomic and molecular phases is obviously 
very different. It is beyond the reach of current and upcoming radio facilities to detect 
\hi\ emission from individual PHIBSS galaxies, but only the presence of large \hi\ reservoirs that 
are unaccounted for (and considered unlikely by current simulations, \eg\ \citealt{lagos14,popping14}) can  make the bulk of PHIBSS 
galaxies significantly different from \highz\ and HIghMass. 

Conversely, DYNAMO galaxies have less overlap with the other three samples considered here. 
Indeed, DYNAMO does not include galaxies with stellar masses above $\sim 10^{11}$ \Msun,
hence the parameter space in common with \highz\ and PHIBSS is somewhat limited. 
Because DYNAMO gas fractions are only estimates, we also plot the four galaxies 
with actual CO measurements (dark green squares, connected by lines to the corresponding KS estimates), 
which agree pretty well with the KS predictions. 
Thus, taking all the predicted gas masses at face value, about half of DYNAMO
systems seem to have gas fractions as extreme as those of the other samples.

Despite their similar total gas fractions, PHIBSS and \highz\ samples are remarkably different when 
we look at their SFRs and stellar mass surface densities in the bottom panels of Fig.~\ref{hz_lit}. 
Firstly, as expected, the SFRs (and hence star formation efficiency) of PHIBSS 
galaxies are significantly higher (\about 1 dex) than all the other samples 
considered here. Secondly, the average stellar mass surface densities of PHIBSS systems
(as well as those of HIghMass and DYNAMO) are entirely consistent with those of disk-dominated, 
local galaxies with similar stellar masses, and only a couple of objects exhibit values 
of \must\ comparable to those of \highz. As mentioned in the previous section, the fact that
\highz\ galaxies occupy a region of parameter space in the \must - \Mst\ diagram that is practically 
untouched by the other samples is the result of a selection effect.
As we are looking at two families of galaxies separated by $\sim$6 Gyr of evolution, 
the differences in optical and SFR properties between \highz\ and PHIBSS samples are not surprising,
and it is interesting to speculate whether or not  we can confidently exclude an evolutionary link between the two. 

If PHIBSS galaxies are representative of the massive population at $z\sim$1, 
it would be extremely hard to explain why typical massive galaxies at $z\sim$1 
evolve into such a rare population of gas-rich systems as \highz. 
Conversely, if gas-rich $z\sim$1 systems are the tip of the iceberg of the high-redshift 
population of disk galaxies, as hinted by recent numerical simulations (\eg\ \citealt{lagos14,popping14}), 
the main obstacle to reconcile the two populations is just the difference 
in \must, which would require significant disk growth from $z\sim$1.2 to $\sim$0.2.
Observations and semi-analytical models \citep{dutton11} suggest that, between 
$z\sim$1 and 0, a massive disk galaxy increases its stellar mass by a factor of 
$\sim$2.5 and its radius by a factor of $\sim$2. This implies a decrease 
in \must\ of $\sim$0.2 dex. Although this shift is too small to 
reconcile \highz\ and PHIBSS galaxies, it is intriguing to note that, once \highz\ 
and HIghMass are treated as a single population of ``local'' \hi\ massive systems, 
the differences with the population of gas-rich turbulent disks at high-redshift 
almost entirely disappear. 

Unfortunately, until a proper characterization of the gas properties 
of a representative population of disk galaxies at $z\sim$1 is obtained, it is impossible 
to establish whether gas-rich galaxies at various redshifts are really linked, and 
this discussion remains just an intriguing speculation. 
Nevertheless, our findings highlight the importance of investigating the still vastly 
unexplored regime of very high gas content for our understanding of galaxy 
evolution across cosmic time.

\section{Summary and conclusions}

In this paper we presented \highz, a survey that measured the \hi\ content of 39 disk galaxies at redshift
$z\sim 0.2$ using the Arecibo radio telescope. This sample includes the highest-redshift detections
of \hi\ emission from individual galaxies published to date, which are also among the most \hi -massive
systems known. By selection, \highz\ galaxies are disk-dominated systems in relatively isolated
fields, with stellar masses \Mst $= 2$-22 $\times 10^{10}$  \Msun, \hi\ masses
\Mhi $= 2$-8 $\times 10^{10}$ \Msun, star formation rates of 3-35 \Msun~yr$^{-1}$
and redshifts $z=0.17$-0.25.

We showed that the \highz\ galaxies have unusually large \hi\ gas fractions and blue \nuvr\ colours 
for their stellar masses. However, when we look at more physical relations, 
such as the gas fraction plane and the baryonic Tully-Fisher relation,
\highz\ galaxies are indistinguishable from the average systems. In other words, their gas content is
exactly what is expected from their UV and optical properties, and there is nothing unusual in the way in 
which star formation proceeds in these galaxies, or in the relation between their dynamical and
baryonic masses. We concluded that \highz\ galaxies are rare, scaled-up 
versions of disk galaxies in the local Universe, and there is no need to invoke unusual episodes
of gas accretion to explain their large reservoirs.

When compared to HIghMass, which includes the most \hi-rich local galaxies extracted from the \hi -blind 
ALFALFA survey, \highz\ systems show striking similarities in their gas content and star formation.
The only significant difference is that, by selection, \highz\ galaxies have \about 0.6 dex lower stellar 
surface densities, suggesting higher values of their spin parameters. Therefore HIghMass galaxies appear 
to be the local counterparts of \highz, with the latter mapping into the high end of the spin parameter 
distribution of HIghMass.

It is more difficult to establish a connection with the gas-rich, turbulent disks identified
by PHIBSS at $z\sim 1$. It is intriguing that the total gas content of \highz\ and PHIBSS seems to be
similar (although the phase is clearly different) but, unless the PHIBSS disks are not representative
of the general disk population at $z\sim 1$, it is very unlikely that they could be the progenitors 
of such a rare population at $z\sim 0.2$.

In addition to probing the \hi\ Universe beyond $z=0.2$ for the first time with direct detections,
the \highz\ survey provides important insights into the properties of the massive, \hi-rich systems
that will likely dominate {\bc the samples detected at higher redshift by future \hi\ surveys} with 
the SKA and its precursor telescopes.
This is particularly relevant for deep \hi\ surveys such as
DINGO \citep[Deep Investigation of Neutral Gas Origins,][]{dingo09} and especially
LADUMA \citep[Looking At the Distant Universe with the MeerKAT Array,][]{laduma12},
which will open the higher redshift Universe to \hi\ exploration.

\section*{Acknowledgments}

We wish to thank Shan Huang for kindly providing us with HIghMass data in advance of 
publication. BC warmly thanks Martha P. Haynes and Riccardo Giovanelli for their
contributions to the initial stages of the survey, and Phil Perillat, Ganesh  
Rajagopalan and the telescope operators at Arecibo for their help and assistance.
We thank our referee, B\"{a}rbel Koribalski, for useful comments that helped us 
improving the clarity of our manuscript.
BC is the recipient of an Australian Research Council Future 
Fellowship (FT120100660). LC acknowledges support under the Australian Research 
Council's Discovery Projects funding scheme (DP130100664).

This research has made use of the NASA/IPAC Extragalactic Database
(NED) which is operated by the Jet Propulsion Laboratory, California
Institute of Technology, under contract with the National Aeronautics
and Space Administration.

The Arecibo Observatory is operated by SRI International under a
cooperative agreement with the National Science Foundation
(AST-1100968), and in alliance with Ana G. M{\'e}ndez-Universidad
Metropolitana, and the Universities Space Research Association.

GALEX (Galaxy Evolution Explorer) is a NASA Small Explorer, launched
in April 2003. We gratefully acknowledge NASA's support for
construction, operation, and science analysis for the GALEX mission,
developed in cooperation with the Centre National d'Etudes Spatiales
(CNES) of France and the Korean Ministry of Science and Technology. 

Funding for the SDSS and SDSS-II has been provided by the Alfred
P. Sloan Foundation, the Participating Institutions, the National
Science Foundation, the U.S. Department of Energy, the National
Aeronautics and Space Administration, the Japanese Monbukagakusho, the
Max Planck Society, and the Higher Education Funding Council for
England. The SDSS Web Site is http://www.sdss.org/.

The SDSS is managed by the Astrophysical Research Consortium for the
Participating Institutions. The Participating Institutions are the
American Museum of Natural History, Astrophysical Institute Potsdam,
University of Basel, University of Cambridge, Case Western Reserve
University, University of Chicago, Drexel University, Fermilab, the
Institute for Advanced Study, the Japan Participation Group, Johns
Hopkins University, the Joint Institute for Nuclear Astrophysics, the
Kavli Institute for Particle Astrophysics and Cosmology, the Korean
Scientist Group, the Chinese Academy of Sciences (LAMOST), Los Alamos
National Laboratory, the Max-Planck-Institute for Astronomy (MPIA),
the Max-Planck-Institute for Astrophysics (MPA), New Mexico State
University, Ohio State University, University of Pittsburgh,
University of Portsmouth, Princeton University, the United States
Naval Observatory, and the University of Washington.


\bibliography{highz}

\begin{thebibliography}{69}
\expandafter\ifx\csname natexlab\endcsname\relax\def\natexlab#1{#1}\fi

\bibitem[{{Abazajian} {et~al}\mbox{.}(2004){Abazajian}, {Adelman-McCarthy},
  {Ag{\"u}eros}, {Allam}, {Anderson}, {Anderson}, {Annis}, {Bahcall}, {Baldry},
  {Bastian}, {Berlind}, {Bernardi}, {Blanton}, {Bochanski}, {Boroski},
  {Briggs}, {Brinkmann}, {Brunner}, {Budav{\'a}ri}, {Carey}, {Carliles},
  {Castander}, {Connolly}, {Csabai}, {Doi}, {Dong}, {Eisenstein}, {Evans},
  {Fan}, {Finkbeiner}, {Friedman}, {Frieman}, {Fukugita}, {Gal}, {Gillespie},
  {Glazebrook}, {Gray}, {Grebel}, {Gunn}, {Gurbani}, {Hall}, {Hamabe},
  {Harris}, {Harris}, {Harvanek}, {Heckman}, {Hendry}, {Hennessy}, {Hindsley},
  {Hogan}, {Hogg}, {Holmgren}, {Ichikawa}, {Ichikawa}, {Ivezi{\'c}}, {Jester},
  {Johnston}, {Jorgensen}, {Kent}, {Kleinman}, {Knapp}, {Kniazev}, {Kron},
  {Krzesinski}, {Kunszt}, {Kuropatkin}, {Lamb}, {Lampeitl}, {Lee}, {Leger},
  {Li}, {Lin}, {Loh}, {Long}, {Loveday}, {Lupton}, {Malik}, {Margon},
  {Matsubara}, {McGehee}, {McKay}, {Meiksin}, {Munn}, {Nakajima}, {Nash},
  {Neilsen}, {Newberg}, {Newman}, {Nichol}, {Nicinski}, {Nieto-Santisteban},
  {Nitta}, {Okamura}, {O'Mullane}, {Ostriker}, {Owen}, {Padmanabhan},
  {Peoples}, {Pier}, {Pope}, {Quinn}, {Richards}, {Richmond}, {Rix}, {Rockosi},
  {Schlegel}, {Schneider}, {Scranton}, {Sekiguchi}, {Seljak}, {Sergey},
  {Sesar}, {Sheldon}, {Shimasaku}, {Siegmund}, {Silvestri}, {Smith}, {Smol{\v
  c}i{\'c}}, {Snedden}, {Stebbins}, {Stoughton}, {Strauss}, {SubbaRao},
  {Szalay}, {Szapudi}, {Szkody}, {Szokoly}, {Tegmark}, {Teodoro}, {Thakar},
  {Tremonti}, {Tucker}, {Uomoto}, {Vanden Berk}, {Vandenberg}, {Vogeley},
  {Voges}, {Vogt}, {Walkowicz}, {Wang}, {Weinberg}, {West}, {White}, {Wilhite},
  {Xu}, {Yanny}, {Yasuda}, {Yip}, {Yocum}, {York}, {Zehavi}, {Zibetti}, \&
  {Zucker}}]{sdss2}
{Abazajian} K. {et~al.}, 2004, \aj, 128, 502

\bibitem[{{Abazajian} {et~al}\mbox{.}(2005){Abazajian}, {Adelman-McCarthy},
  {Ag{\"u}eros}, {Allam}, {Anderson}, {Anderson}, {Annis}, {Bahcall}, {Baldry},
  {Bastian}, {Berlind}, {Bernardi}, {Blanton}, {Bochanski}, {Boroski},
  {Brewington}, {Briggs}, {Brinkmann}, {Brunner}, {Budav{\'a}ri}, {Carey},
  {Castander}, {Connolly}, {Covey}, {Csabai}, {Dalcanton}, {Doi}, {Dong},
  {Eisenstein}, {Evans}, {Fan}, {Finkbeiner}, {Friedman}, {Frieman},
  {Fukugita}, {Gillespie}, {Glazebrook}, {Gray}, {Grebel}, {Gunn}, {Gurbani},
  {Hall}, {Hamabe}, {Harbeck}, {Harris}, {Harris}, {Harvanek}, {Hawley},
  {Hayes}, {Heckman}, {Hendry}, {Hennessy}, {Hindsley}, {Hogan}, {Hogg},
  {Holmgren}, {Holtzman}, {Ichikawa}, {Ichikawa}, {Ivezi{\'c}}, {Jester},
  {Johnston}, {Jorgensen}, {Juri{\'c}}, {Kent}, {Kleinman}, {Knapp}, {Kniazev},
  {Kron}, {Krzesinski}, {Lamb}, {Lampeitl}, {Lee}, {Lin}, {Long}, {Loveday},
  {Lupton}, {Mannery}, {Margon}, {Mart{\'{\i}}nez-Delgado}, {Matsubara},
  {McGehee}, {McKay}, {Meiksin}, {M{\'e}nard}, {Munn}, {Nash}, {Neilsen},
  {Newberg}, {Newman}, {Nichol}, {Nicinski}, {Nieto-Santisteban}, {Nitta},
  {Okamura}, {O'Mullane}, {Owen}, {Padmanabhan}, {Pauls}, {Peoples}, {Pier},
  {Pope}, {Pourbaix}, {Quinn}, {Raddick}, {Richards}, {Richmond}, {Rix},
  {Rockosi}, {Schlegel}, {Schneider}, {Schroeder}, {Scranton}, {Sekiguchi},
  {Sheldon}, {Shimasaku}, {Silvestri}, {Smith}, {Smol{\v c}i{\'c}}, {Snedden},
  {Stebbins}, {Stoughton}, {Strauss}, {SubbaRao}, {Szalay}, {Szapudi},
  {Szkody}, {Szokoly}, {Tegmark}, {Teodoro}, {Thakar}, {Tremonti}, {Tucker},
  {Uomoto}, {Vanden Berk}, {Vandenberg}, {Vogeley}, {Voges}, {Vogt},
  {Walkowicz}, {Wang}, {Weinberg}, {West}, {White}, {Wilhite}, {Xu}, {Yanny},
  {Yasuda}, {Yip}, {Yocum}, {York}, {Zehavi}, {Zibetti}, \& {Zucker}}]{sdss3}
{Abazajian} K. {et~al.}, 2005, \aj, 129, 1755

\bibitem[{{Abazajian} {et~al}\mbox{.}(2003){Abazajian}, {Adelman-McCarthy},
  {Ag{\"u}eros}, {Allam}, {Anderson}, {Annis}, {Bahcall}, {Baldry}, {Bastian},
  {Berlind}, {Bernardi}, {Blanton}, {Blythe}, {Bochanski}, {Boroski},
  {Brewington}, {Briggs}, {Brinkmann}, {Brunner}, {Budav{\'a}ri}, {Carey},
  {Carr}, {Castander}, {Chiu}, {Collinge}, {Connolly}, {Covey}, {Csabai},
  {Dalcanton}, {Dodelson}, {Doi}, {Dong}, {Eisenstein}, {Evans}, {Fan},
  {Feldman}, {Finkbeiner}, {Friedman}, {Frieman}, {Fukugita}, {Gal},
  {Gillespie}, {Glazebrook}, {Gonzalez}, {Gray}, {Grebel}, {Grodnicki}, {Gunn},
  {Gurbani}, {Hall}, {Hao}, {Harbeck}, {Harris}, {Harris}, {Harvanek},
  {Hawley}, {Heckman}, {Helmboldt}, {Hendry}, {Hennessy}, {Hindsley}, {Hogg},
  {Holmgren}, {Holtzman}, {Homer}, {Hui}, {Ichikawa}, {Ichikawa}, {Inkmann},
  {Ivezi{\'c}}, {Jester}, {Johnston}, {Jordan}, {Jordan}, {Jorgensen},
  {Juri{\'c}}, {Kauffmann}, {Kent}, {Kleinman}, {Knapp}, {Kniazev}, {Kron},
  {Krzesi{\'n}ski}, {Kunszt}, {Kuropatkin}, {Lamb}, {Lampeitl}, {Laubscher},
  {Lee}, {Leger}, {Li}, {Lidz}, {Lin}, {Loh}, {Long}, {Loveday}, {Lupton},
  {Malik}, {Margon}, {McGehee}, {McKay}, {Meiksin}, {Miknaitis}, {Moorthy},
  {Munn}, {Murphy}, {Nakajima}, {Narayanan}, {Nash}, {Neilsen}, {Newberg},
  {Newman}, {Nichol}, {Nicinski}, {Nieto-Santisteban}, {Nitta}, {Odenkirchen},
  {Okamura}, {Ostriker}, {Owen}, {Padmanabhan}, {Peoples}, {Pier}, {Pindor},
  {Pope}, {Quinn}, {Rafikov}, {Raymond}, {Richards}, {Richmond}, {Rix},
  {Rockosi}, {Schaye}, {Schlegel}, {Schneider}, {Schroeder}, {Scranton},
  {Sekiguchi}, {Seljak}, {Sergey}, {Sesar}, {Sheldon}, {Shimasaku}, {Siegmund},
  {Silvestri}, {Sinisgalli}, {Sirko}, {Smith}, {Smol{\v c}i{\'c}}, {Snedden},
  {Stebbins}, {Steinhardt}, {Stinson}, {Stoughton}, {Strateva}, {Strauss},
  {SubbaRao}, {Szalay}, {Szapudi}, {Szkody}, {Tasca}, {Tegmark}, {Thakar},
  {Tremonti}, {Tucker}, {Uomoto}, {Vanden Berk}, {Vandenberg}, {Vogeley},
  {Voges}, {Vogt}, {Walkowicz}, {Weinberg}, {West}, {White}, {Wilhite},
  {Willman}, {Xu}, {Yanny}, {Yarger}, {Yasuda}, {Yip}, {Yocum}, {York},
  {Zakamska}, {Zehavi}, {Zheng}, {Zibetti}, \& {Zucker}}]{sdss1}
{Abazajian} K. {et~al.}, 2003, \aj, 126, 2081

\bibitem[{{Abazajian} {et~al}\mbox{.}(2009){Abazajian}, {Adelman-McCarthy},
  {Ag{\"u}eros}, {Allam}, {Allende Prieto}, {An}, {Anderson}, {Anderson},
  {Annis}, {Bahcall}, \& et~al.}]{sdss7}
{Abazajian} K.~N. {et~al.}, 2009, \apjs, 182, 543

\bibitem[{{Adelman-McCarthy} {et~al}\mbox{.}(2007){Adelman-McCarthy},
  {Ag{\"u}eros}, {Allam}, {Anderson}, {Anderson}, {Annis}, {Bahcall},
  {Bailer-Jones}, {Baldry}, {Barentine}, {Beers}, {Belokurov}, {Berlind},
  {Bernardi}, {Blanton}, {Bochanski}, {Boroski}, {Bramich}, {Brewington},
  {Brinchmann}, {Brinkmann}, {Brunner}, {Budav{\'a}ri}, {Carey}, {Carliles},
  {Carr}, {Castander}, {Connolly}, {Cool}, {Cunha}, {Csabai}, {Dalcanton},
  {Doi}, {Eisenstein}, {Evans}, {Evans}, {Fan}, {Finkbeiner}, {Friedman},
  {Frieman}, {Fukugita}, {Gillespie}, {Gilmore}, {Glazebrook}, {Gray},
  {Grebel}, {Gunn}, {de Haas}, {Hall}, {Harvanek}, {Hawley}, {Hayes},
  {Heckman}, {Hendry}, {Hennessy}, {Hindsley}, {Hirata}, {Hogan}, {Hogg},
  {Holtzman}, {Ichikawa}, {Ichikawa}, {Ivezi{\'c}}, {Jester}, {Johnston},
  {Jorgensen}, {Juri{\'c}}, {Kauffmann}, {Kent}, {Kleinman}, {Knapp},
  {Kniazev}, {Kron}, {Krzesinski}, {Kuropatkin}, {Lamb}, {Lampeitl}, {Lee},
  {Leger}, {Lima}, {Lin}, {Long}, {Loveday}, {Lupton}, {Mandelbaum}, {Margon},
  {Mart{\'{\i}}nez-Delgado}, {Matsubara}, {McGehee}, {McKay}, {Meiksin},
  {Munn}, {Nakajima}, {Nash}, {Neilsen}, {Newberg}, {Nichol},
  {Nieto-Santisteban}, {Nitta}, {Oyaizu}, {Okamura}, {Ostriker}, {Padmanabhan},
  {Park}, {Peoples}, {Pier}, {Pope}, {Pourbaix}, {Quinn}, {Raddick}, {Re
  Fiorentin}, {Richards}, {Richmond}, {Rix}, {Rockosi}, {Schlegel},
  {Schneider}, {Scranton}, {Seljak}, {Sheldon}, {Shimasaku}, {Silvestri},
  {Smith}, {Smol{\v c}i{\'c}}, {Snedden}, {Stebbins}, {Stoughton}, {Strauss},
  {SubbaRao}, {Suto}, {Szalay}, {Szapudi}, {Szkody}, {Tegmark}, {Thakar},
  {Tremonti}, {Tucker}, {Uomoto}, {Vanden Berk}, {Vandenberg}, {Vidrih},
  {Vogeley}, {Voges}, {Vogt}, {Weinberg}, {West}, {White}, {Wilhite}, {Yanny},
  {Yocum}, {York}, {Zehavi}, {Zibetti}, \& {Zucker}}]{sdss5}
{Adelman-McCarthy} J.~K. {et~al.}, 2007, \apjs, 172, 634

\bibitem[{{Barnes} {et~al}\mbox{.}(2001){Barnes}, {Staveley-Smith}, {de Blok},
  {Oosterloo}, {Stewart}, {Wright}, {Banks}, {Bhathal}, {Boyce}, {Calabretta},
  {Disney}, {Drinkwater}, {Ekers}, {Freeman}, {Gibson}, {Green}, {Haynes}, {te
  Lintel Hekkert}, {Henning}, {Jerjen}, {Juraszek}, {Kesteven}, {Kilborn},
  {Knezek}, {Koribalski}, {Kraan-Korteweg}, {Malin}, {Marquarding}, {Minchin},
  {Mould}, {Price}, {Putman}, {Ryder}, {Sadler}, {Schr{\"o}der}, {Stootman},
  {Webster}, {Wilson}, \& {Ye}}]{hipass}
{Barnes} D.~G. {et~al.}, 2001, \mnras, 322, 486

\bibitem[{{Bell} {et~al}\mbox{.}(2005){Bell}, {Papovich}, {Wolf}, {Le Floc'h},
  {Caldwell}, {Barden}, {Egami}, {McIntosh}, {Meisenheimer},
  {P{\'e}rez-Gonz{\'a}lez}, {Rieke}, {Rieke}, {Rigby}, \& {Rix}}]{bell05}
{Bell} E.~F. {et~al.}, 2005, \apj, 625, 23

\bibitem[{{Boissier} \& {Prantzos}(2000)}]{boissier00}
{Boissier} S., {Prantzos} N., 2000, \mnras, 312, 398

\bibitem[{{Booth} {et~al}\mbox{.}(2009){Booth}, {de Blok}, {Jonas}, \&
  {Fanaroff}}]{meerkat}
{Booth} R.~S., {de Blok} W.~J.~G., {Jonas} J.~L., {Fanaroff} B., 2009, ArXiv
  e-prints

\bibitem[{{Boselli} {et~al}\mbox{.}(2014){Boselli}, {Cortese}, {Boquien},
  {Boissier}, {Catinella}, {Lagos}, \& {Saintonge}}]{boselli14b}
{Boselli} A., {Cortese} L., {Boquien} M., {Boissier} S., {Catinella} B.,
  {Lagos} C., {Saintonge} A., 2014, \aap, 564, A66

\bibitem[{{Bothun} {et~al}\mbox{.}(1987){Bothun}, {Impey}, {Malin}, \&
  {Mould}}]{bothun87}
{Bothun} G.~D., {Impey} C.~D., {Malin} D.~F., {Mould} J.~R., 1987, \aj, 94, 23

\bibitem[{{Brinchmann} {et~al}\mbox{.}(2004){Brinchmann}, {Charlot}, {White},
  {Tremonti}, {Kauffmann}, {Heckman}, \& {Brinkmann}}]{jarle04}
{Brinchmann} J., {Charlot} S., {White} S.~D.~M., {Tremonti} C., {Kauffmann} G.,
  {Heckman} T., {Brinkmann} J., 2004, \mnras, 351, 1151

\bibitem[{{Carilli} \& {Rawlings}(2004)}]{ska}
{Carilli} C.~L., {Rawlings} S., 2004, \nar, 48, 979

\bibitem[{{Catinella}, {Haynes} \& {Giovanelli}(2007){Catinella}, {Haynes}, \&
  {Giovanelli}}]{widths}
{Catinella} B., {Haynes} M.~P., {Giovanelli} R., 2007, \aj, 134, 334

\bibitem[{{Catinella} {et~al}\mbox{.}(2008){Catinella}, {Haynes}, {Giovanelli},
  {Gardner}, \& {Connolly}}]{highz_pilot}
{Catinella} B., {Haynes} M.~P., {Giovanelli} R., {Gardner} J.~P., {Connolly}
  A.~J., 2008, \apjl, 685, L13

\bibitem[{{Catinella} {et~al}\mbox{.}(2012{\natexlab{a}}){Catinella},
  {Kauffmann}, {Schiminovich}, {Lemonias}, {Scannapieco}, {Wang}, {Fabello},
  {Hummels}, {Moran}, {Wu}, {Cooper}, {Giovanelli}, {Haynes}, {Heckman}, \&
  {Saintonge}}]{gass4}
{Catinella} B. {et~al.}, 2012{\natexlab{a}}, \mnras, 420, 1959

\bibitem[{{Catinella} {et~al}\mbox{.}(2013){Catinella}, {Schiminovich},
  {Cortese}, {Fabello}, {Hummels}, {Moran}, {Lemonias}, {Cooper}, {Wu},
  {Heckman}, \& {Wang}}]{gass_dr3}
{Catinella} B. {et~al.}, 2013, \mnras, 436, 34

\bibitem[{{Catinella} {et~al}\mbox{.}(2012{\natexlab{b}}){Catinella},
  {Schiminovich}, {Kauffmann}, {Fabello}, {Hummels}, {Lemonias}, {Moran}, {Wu},
  {Cooper}, \& {Wang}}]{gass_dr2}
{Catinella} B. {et~al.}, 2012{\natexlab{b}}, \aap, 544, A65

\bibitem[{{Catinella} {et~al}\mbox{.}(2010){Catinella}, {Schiminovich},
  {Kauffmann}, {Fabello}, {Wang}, {Hummels}, {Lemonias}, {Moran}, {Wu},
  {Giovanelli}, {Haynes}, {Heckman}, {Basu-Zych}, {Blanton}, {Brinchmann},
  {Budav{\'a}ri}, {Gon{\c c}alves}, {Johnson}, {Kennicutt}, {Madore}, {Martin},
  {Rich}, {Tacconi}, {Thilker}, {Wild}, \& {Wyder}}]{gass1}
{Catinella} B. {et~al.}, 2010, \mnras, 403, 683

\bibitem[{{Chabrier}(2003)}]{chabrier03}
{Chabrier} G., 2003, \pasp, 115, 763

\bibitem[{{Chilingarian}, {Melchior} \& {Zolotukhin}(2010){Chilingarian},
  {Melchior}, \& {Zolotukhin}}]{chilingarian10}
{Chilingarian} I.~V., {Melchior} A.-L., {Zolotukhin} I.~Y., 2010, \mnras, 405,
  1409

\bibitem[{{Cluver} {et~al}\mbox{.}(2010){Cluver}, {Jarrett}, {Kraan-Korteweg},
  {Koribalski}, {Appleton}, {Melbourne}, {Emonts}, \& {Woudt}}]{cluver10}
{Cluver} M.~E., {Jarrett} T.~H., {Kraan-Korteweg} R.~C., {Koribalski} B.~S.,
  {Appleton} P.~N., {Melbourne} J., {Emonts} B., {Woudt} P.~A., 2010, \apj,
  725, 1550

\bibitem[{{Cortese}(2012)}]{luca12_redsp}
{Cortese} L., 2012, \aap, 543, A132

\bibitem[{{Cortese} {et~al}\mbox{.}(2011){Cortese}, {Catinella}, {Boissier},
  {Boselli}, \& {Heinis}}]{luca11}
{Cortese} L., {Catinella} B., {Boissier} S., {Boselli} A., {Heinis} S., 2011,
  \mnras, 415, 1797

\bibitem[{{Daddi} {et~al}\mbox{.}(2010){Daddi}, {Bournaud}, {Walter},
  {Dannerbauer}, {Carilli}, {Dickinson}, {Elbaz}, {Morrison}, {Riechers},
  {Onodera}, {Salmi}, {Krips}, \& {Stern}}]{daddi10}
{Daddi} E. {et~al.}, 2010, \apj, 713, 686

\bibitem[{{Dutton} {et~al}\mbox{.}(2011){Dutton}, {van den Bosch}, {Faber},
  {Simard}, {Kassin}, {Koo}, {Bundy}, {Huang}, {Weiner}, {Cooper}, {Newman},
  {Mozena}, \& {Koekemoer}}]{dutton11}
{Dutton} A.~A. {et~al.}, 2011, \mnras, 410, 1660

\bibitem[{{Fabello} {et~al}\mbox{.}(2011){Fabello}, {Catinella}, {Giovanelli},
  {Kauffmann}, {Haynes}, {Heckman}, \& {Schiminovich}}]{fabello1}
{Fabello} S., {Catinella} B., {Giovanelli} R., {Kauffmann} G., {Haynes} M.~P.,
  {Heckman} T.~M., {Schiminovich} D., 2011, \mnras, 411, 993

\bibitem[{{Fern{\'a}ndez} {et~al}\mbox{.}(2013){Fern{\'a}ndez}, {van Gorkom},
  {Hess}, {Pisano}, {Kreckel}, {Momjian}, {Popping}, {Oosterloo}, {Chomiuk},
  {Verheijen}, {Henning}, {Schiminovich}, {Bershady}, {Wilcots}, \&
  {Scoville}}]{chiles_pilot}
{Fern{\'a}ndez} X. {et~al.}, 2013, \apjl, 770, L29

\bibitem[{{Fisher} {et~al}\mbox{.}(2014){Fisher}, {Glazebrook}, {Bolatto},
  {Obreschkow}, {Mentuch Cooper}, {Wisnioski}, {Bassett}, {Abraham},
  {Damjanov}, {Green}, \& {McGregor}}]{fisher14}
{Fisher} D.~B. {et~al.}, 2014, \apjl, 790, L30

\bibitem[{{Fridman} \& {Baan}(2001)}]{fridman01}
{Fridman} P.~A., {Baan} W.~A., 2001, \aap, 378, 327

\bibitem[{{Genzel} {et~al}\mbox{.}(2011){Genzel}, {Newman}, {Jones},
  {F{\"o}rster Schreiber}, {Shapiro}, {Genel}, {Lilly}, {Renzini}, {Tacconi},
  {Bouch{\'e}}, {Burkert}, {Cresci}, {Buschkamp}, {Carollo}, {Ceverino},
  {Davies}, {Dekel}, {Eisenhauer}, {Hicks}, {Kurk}, {Lutz}, {Mancini}, {Naab},
  {Peng}, {Sternberg}, {Vergani}, \& {Zamorani}}]{genzel11}
{Genzel} R. {et~al.}, 2011, \apj, 733, 101

\bibitem[{{Giovanelli} {et~al}\mbox{.}(2005){Giovanelli}, {Haynes}, {Kent},
  {Perillat}, {Saintonge}, {Brosch}, {Catinella}, {Hoffman}, {Stierwalt},
  {Spekkens}, {Lerner}, {Masters}, {Momjian}, {Rosenberg}, {Springob},
  {Boselli}, {Charmandaris}, {Darling}, {Davies}, {Lambas}, {Gavazzi},
  {Giovanardi}, {Hardy}, {Hunt}, {Iovino}, {Karachentsev}, {Karachentseva},
  {Koopmann}, {Marinoni}, {Minchin}, {Muller}, {Putman}, {Pantoja}, {Salzer},
  {Scodeggio}, {Skillman}, {Solanes}, {Valotto}, {van Driel}, \& {van
  Zee}}]{alfalfa}
{Giovanelli} R. {et~al.}, 2005, \aj, 130, 2598

\bibitem[{{Green} {et~al}\mbox{.}(2014){Green}, {Glazebrook}, {McGregor},
  {Damjanov}, {Wisnioski}, {Abraham}, {Colless}, {Sharp}, {Crain}, {Poole}, \&
  {McCarthy}}]{dynamo}
{Green} A.~W. {et~al.}, 2014, \mnras, 437, 1070

\bibitem[{{Haynes} {et~al}\mbox{.}(2011){Haynes}, {Giovanelli}, {Martin},
  {Hess}, {Saintonge}, {Adams}, {Hallenbeck}, {Hoffman}, {Huang}, {Kent},
  {Koopmann}, {Papastergis}, {Stierwalt}, {Balonek}, {Craig}, {Higdon},
  {Kornreich}, {Miller}, {O'Donoghue}, {Olowin}, {Rosenberg}, {Spekkens},
  {Troischt}, \& {Wilcots}}]{alfalfa40}
{Haynes} M.~P. {et~al.}, 2011, \aj, 142, 170

\bibitem[{{Holwerda}, {Blyth} \& {Baker}(2012){Holwerda}, {Blyth}, \&
  {Baker}}]{laduma12}
{Holwerda} B.~W., {Blyth} S.-L., {Baker} A.~J., 2012, in IAU Symposium, Vol.
  284, IAU Symposium, {Tuffs} R.~J., {Popescu} C.~C., eds., pp. 496--499

\bibitem[{{Huang} {et~al}\mbox{.}(2012){Huang}, {Haynes}, {Giovanelli}, \&
  {Brinchmann}}]{huang12}
{Huang} S., {Haynes} M.~P., {Giovanelli} R., {Brinchmann} J., 2012, \apj, 756,
  113

\bibitem[{{Huang} {et~al}\mbox{.}(2014){Huang}, {Haynes}, {Giovanelli},
  {Hallenbeck}, {Jones}, {Adams}, {Brinchmann}, {Chengalur}, {Hunt}, {Masters},
  {Matsushita}, {Saintonge}, \& {Spekkens}}]{HIghMass}
{Huang} S. {et~al.}, 2014, \apj, 793, 40

\bibitem[{{Impey} \& {Bothun}(1989)}]{impey89}
{Impey} C., {Bothun} G., 1989, \apj, 341, 89

\bibitem[{{Jaff{\'e}} {et~al}\mbox{.}(2012){Jaff{\'e}}, {Poggianti},
  {Verheijen}, {Deshev}, \& {van Gorkom}}]{jaffe12}
{Jaff{\'e}} Y.~L., {Poggianti} B.~M., {Verheijen} M.~A.~W., {Deshev} B.~Z.,
  {van Gorkom} J.~H., 2012, \apjl, 756, L28

\bibitem[{{Johnston} {et~al}\mbox{.}(2008){Johnston}, {Taylor}, {Bailes},
  {Bartel}, {Baugh}, {Bietenholz}, {Blake}, {Braun}, {Brown}, {Chatterjee},
  {Darling}, {Deller}, {Dodson}, {Edwards}, {Ekers}, {Ellingsen}, {Feain},
  {Gaensler}, {Haverkorn}, {Hobbs}, {Hopkins}, {Jackson}, {James}, {Joncas},
  {Kaspi}, {Kilborn}, {Koribalski}, {Kothes}, {Landecker}, {Lenc}, {Lovell},
  {Macquart}, {Manchester}, {Matthews}, {McClure-Griffiths}, {Norris}, {Pen},
  {Phillips}, {Power}, {Protheroe}, {Sadler}, {Schmidt}, {Stairs},
  {Staveley-Smith}, {Stil}, {Tingay}, {Tzioumis}, {Walker}, {Wall}, \&
  {Wolleben}}]{askap}
{Johnston} S. {et~al.}, 2008, Experimental Astronomy, 22, 151

\bibitem[{{Kennicutt}(1998)}]{kennicutt98}
{Kennicutt}, Jr. R.~C., 1998, \apj, 498, 541

\bibitem[{{Lagos} {et~al}\mbox{.}(2011){Lagos}, {Baugh}, {Lacey}, {Benson},
  {Kim}, \& {Power}}]{lagos11}
{Lagos} C.~D.~P., {Baugh} C.~M., {Lacey} C.~G., {Benson} A.~J., {Kim} H.-S.,
  {Power} C., 2011, \mnras, 418, 1649

\bibitem[{{Lagos} {et~al}\mbox{.}(2014){Lagos}, {Baugh}, {Zwaan}, {Lacey},
  {Gonzalez-Perez}, {Power}, {Swinbank}, \& {van Kampen}}]{lagos14}
{Lagos} C.~D.~P., {Baugh} C.~M., {Zwaan} M.~A., {Lacey} C.~G., {Gonzalez-Perez}
  V., {Power} C., {Swinbank} A.~M., {van Kampen} E., 2014, \mnras, 440, 920

\bibitem[{{Lah} {et~al}\mbox{.}(2009){Lah}, {Pracy}, {Chengalur}, {Briggs},
  {Colless}, {de Propris}, {Ferris}, {Schmidt}, \& {Tucker}}]{lah09}
{Lah} P. {et~al.}, 2009, \mnras, 399, 1447

\bibitem[{{Lee} {et~al}\mbox{.}(2014){Lee}, {Chung}, {Yun}, {Cybulski},
  {Narayanan}, \& {Erickson}}]{HImonsters}
{Lee} C., {Chung} A., {Yun} M.~S., {Cybulski} R., {Narayanan} G., {Erickson}
  N., 2014, \mnras, 441, 1363

\bibitem[{{Lelli}, {Fraternali} \& {Sancisi}(2010){Lelli}, {Fraternali}, \&
  {Sancisi}}]{lelli10}
{Lelli} F., {Fraternali} F., {Sancisi} R., 2010, \aap, 516, A11

\bibitem[{{Lilly} {et~al}\mbox{.}(1996){Lilly}, {Le Fevre}, {Hammer}, \&
  {Crampton}}]{lilly96}
{Lilly} S.~J., {Le Fevre} O., {Hammer} F., {Crampton} D., 1996, \apjl, 460, L1+

\bibitem[{{Madau} {et~al}\mbox{.}(1996){Madau}, {Ferguson}, {Dickinson},
  {Giavalisco}, {Steidel}, \& {Fruchter}}]{madau96}
{Madau} P., {Ferguson} H.~C., {Dickinson} M.~E., {Giavalisco} M., {Steidel}
  C.~C., {Fruchter} A., 1996, \mnras, 283, 1388

\bibitem[{{Martin} {et~al}\mbox{.}(2010){Martin}, {Papastergis}, {Giovanelli},
  {Haynes}, {Springob}, \& {Stierwalt}}]{himf_aa}
{Martin} A.~M., {Papastergis} E., {Giovanelli} R., {Haynes} M.~P., {Springob}
  C.~M., {Stierwalt} S., 2010, \apj, 723, 1359

\bibitem[{{Martin} {et~al}\mbox{.}(2005){Martin}, {Fanson}, {Schiminovich},
  {Morrissey}, {Friedman}, {Barlow}, {Conrow}, {Grange}, {Jelinsky},
  {Milliard}, {Siegmund}, {Bianchi}, {Byun}, {Donas}, {Forster}, {Heckman},
  {Lee}, {Madore}, {Malina}, {Neff}, {Rich}, {Small}, {Surber}, {Szalay},
  {Welsh}, \& {Wyder}}]{galex}
{Martin} D.~C. {et~al.}, 2005, \apjl, 619, L1

\bibitem[{{McGaugh} {et~al}\mbox{.}(2000){McGaugh}, {Schombert}, {Bothun}, \&
  {de Blok}}]{mcgaugh00}
{McGaugh} S.~S., {Schombert} J.~M., {Bothun} G.~D., {de Blok} W.~J.~G., 2000,
  \apjl, 533, L99

\bibitem[{{Meyer}(2009)}]{dingo09}
{Meyer} M., 2009, in Panoramic Radio Astronomy: Wide-field 1-2 GHz Research on
  Galaxy Evolution

\bibitem[{{Mo}, {Mao} \& {White}(1998){Mo}, {Mao}, \& {White}}]{mmw98}
{Mo} H.~J., {Mao} S., {White} S.~D.~M., 1998, \mnras, 295, 319

\bibitem[{{Obreschkow} \& {Rawlings}(2009)}]{obreschkow09}
{Obreschkow} D., {Rawlings} S., 2009, \apjl, 696, L129

\bibitem[{{Papastergis} {et~al}\mbox{.}(2013){Papastergis}, {Giovanelli},
  {Haynes}, {Rodr{\'{\i}}guez-Puebla}, \& {Jones}}]{papastergis13}
{Papastergis} E., {Giovanelli} R., {Haynes} M.~P., {Rodr{\'{\i}}guez-Puebla}
  A., {Jones} M.~G., 2013, \apj, 776, 43

\bibitem[{{Pickering} {et~al}\mbox{.}(1997){Pickering}, {Impey}, {van Gorkom},
  \& {Bothun}}]{pickering97}
{Pickering} T.~E., {Impey} C.~D., {van Gorkom} J.~H., {Bothun} G.~D., 1997,
  \aj, 114, 1858

\bibitem[{{Popping}, {Somerville} \& {Trager}(2014){Popping}, {Somerville}, \&
  {Trager}}]{popping14}
{Popping} G., {Somerville} R.~S., {Trager} S.~C., 2014, \mnras, 442, 2398

\bibitem[{{Saintonge} {et~al}\mbox{.}(2011){Saintonge}, {Kauffmann}, {Kramer},
  {Tacconi}, {Buchbender}, {Catinella}, {Fabello}, {Graci{\'a}-Carpio}, {Wang},
  {Cortese}, {Fu}, {Genzel}, {Giovanelli}, {Guo}, {Haynes}, {Heckman},
  {Krumholz}, {Lemonias}, {Li}, {Moran}, {Rodriguez-Fernandez}, {Schiminovich},
  {Schuster}, \& {Sievers}}]{coldgass1}
{Saintonge} A. {et~al.}, 2011, \mnras, 415, 32

\bibitem[{{Salim} {et~al}\mbox{.}(2007){Salim}, {Rich}, {Charlot},
  {Brinchmann}, {Johnson}, {Schiminovich}, {Seibert}, {Mallery}, {Heckman},
  {Forster}, {Friedman}, {Martin}, {Morrissey}, {Neff}, {Small}, {Wyder},
  {Bianchi}, {Donas}, {Lee}, {Madore}, {Milliard}, {Szalay}, {Welsh}, \&
  {Yi}}]{salim07}
{Salim} S. {et~al.}, 2007, \apjs, 173, 267

\bibitem[{{Schiminovich} {et~al}\mbox{.}(2010){Schiminovich}, {Catinella},
  {Kauffmann}, {Fabello}, {Wang}, {Hummels}, {Lemonias}, {Moran}, {Wu},
  {Giovanelli}, {Haynes}, {Heckman}, {Basu-Zych}, {Blanton}, {Brinchmann},
  {Budav{\'a}ri}, {Gon{\c c}alves}, {Johnson}, {Kennicutt}, {Madore}, {Martin},
  {Rich}, {Tacconi}, {Thilker}, {Wild}, \& {Wyder}}]{gass2}
{Schiminovich} D. {et~al.}, 2010, \mnras, 408, 919

\bibitem[{{Seibert} {et~al}\mbox{.}(2012){Seibert}, {Wyder}, {Neill}, {Madore},
  {Bianchi}, {Smith}, {Shiao}, {Schiminovich}, {Rich}, {Conrow}, {Martin}, \&
  {GALEX Catalog Team}}]{seibert12}
{Seibert} M. {et~al.}, 2012, in American Astronomical Society Meeting
  Abstracts, Vol. 219, American Astronomical Society Meeting Abstracts \#219,
  p. 340.01

\bibitem[{{Springob} {et~al}\mbox{.}(2005){Springob}, {Haynes}, {Giovanelli},
  \& {Kent}}]{springob05}
{Springob} C.~M., {Haynes} M.~P., {Giovanelli} R., {Kent} B.~R., 2005, \apjs,
  160, 149

\bibitem[{{Tacconi} {et~al}\mbox{.}(2013){Tacconi}, {Neri}, {Genzel}, {Combes},
  {Bolatto}, {Cooper}, {Wuyts}, {Bournaud}, {Burkert}, {Comerford}, {Cox},
  {Davis}, {F{\"o}rster Schreiber}, {Garc{\'{\i}}a-Burillo}, {Gracia-Carpio},
  {Lutz}, {Naab}, {Newman}, {Omont}, {Saintonge}, {Shapiro Griffin}, {Shapley},
  {Sternberg}, \& {Weiner}}]{phibss}
{Tacconi} L.~J. {et~al.}, 2013, \apj, 768, 74

\bibitem[{{van Zee} {et~al}\mbox{.}(1997){van Zee}, {Maddalena}, {Haynes},
  {Hogg}, \& {Roberts}}]{vanzee97}
{van Zee} L., {Maddalena} R.~J., {Haynes} M.~P., {Hogg} D.~E., {Roberts} M.~S.,
  1997, \aj, 113, 1638

\bibitem[{{Verheijen} {et~al}\mbox{.}(2007){Verheijen}, {van Gorkom},
  {Szomoru}, {Dwarakanath}, {Poggianti}, \& {Schiminovich}}]{verheijen07}
{Verheijen} M., {van Gorkom} J.~H., {Szomoru} A., {Dwarakanath} K.~S.,
  {Poggianti} B.~M., {Schiminovich} D., 2007, \apjl, 668, L9

\bibitem[{{Wang} {et~al}\mbox{.}(2013){Wang}, {Kauffmann}, {J{\'o}zsa},
  {Serra}, {van der Hulst}, {Bigiel}, {Brinchmann}, {Verheijen}, {Oosterloo},
  {Wang}, {Li}, {den Heijer}, \& {Kerp}}]{bluedisks}
{Wang} J. {et~al.}, 2013, \mnras, 433, 270

\bibitem[{{Wyder} {et~al}\mbox{.}(2007){Wyder}, {Martin}, {Schiminovich},
  {Seibert}, {Budav{\'a}ri}, {Treyer}, {Barlow}, {Forster}, {Friedman},
  {Morrissey}, {Neff}, {Small}, {Bianchi}, {Donas}, {Heckman}, {Lee}, {Madore},
  {Milliard}, {Rich}, {Szalay}, {Welsh}, \& {Yi}}]{wyder07}
{Wyder} T.~K. {et~al.}, 2007, \apjs, 173, 293

\bibitem[{{York} {et~al}\mbox{.}(2000){York}, {Adelman}, {Anderson},
  {Anderson}, {Annis}, {Bahcall}, {Bakken}, {Barkhouser}, {Bastian}, {Berman},
  {Boroski}, {Bracker}, {Briegel}, {Briggs}, {Brinkmann}, {Brunner}, {Burles},
  {Carey}, {Carr}, {Castander}, {Chen}, {Colestock}, {Connolly}, {Crocker},
  {Csabai}, {Czarapata}, {Davis}, {Doi}, {Dombeck}, {Eisenstein}, {Ellman},
  {Elms}, {Evans}, {Fan}, {Federwitz}, {Fiscelli}, {Friedman}, {Frieman},
  {Fukugita}, {Gillespie}, {Gunn}, {Gurbani}, {de Haas}, {Haldeman}, {Harris},
  {Hayes}, {Heckman}, {Hennessy}, {Hindsley}, {Holm}, {Holmgren}, {Huang},
  {Hull}, {Husby}, {Ichikawa}, {Ichikawa}, {Ivezi{\'c}}, {Kent}, {Kim},
  {Kinney}, {Klaene}, {Kleinman}, {Kleinman}, {Knapp}, {Korienek}, {Kron},
  {Kunszt}, {Lamb}, {Lee}, {Leger}, {Limmongkol}, {Lindenmeyer}, {Long},
  {Loomis}, {Loveday}, {Lucinio}, {Lupton}, {MacKinnon}, {Mannery}, {Mantsch},
  {Margon}, {McGehee}, {McKay}, {Meiksin}, {Merelli}, {Monet}, {Munn},
  {Narayanan}, {Nash}, {Neilsen}, {Neswold}, {Newberg}, {Nichol}, {Nicinski},
  {Nonino}, {Okada}, {Okamura}, {Ostriker}, {Owen}, {Pauls}, {Peoples},
  {Peterson}, {Petravick}, {Pier}, {Pope}, {Pordes}, {Prosapio},
  {Rechenmacher}, {Quinn}, {Richards}, {Richmond}, {Rivetta}, {Rockosi},
  {Ruthmansdorfer}, {Sandford}, {Schlegel}, {Schneider}, {Sekiguchi}, {Sergey},
  {Shimasaku}, {Siegmund}, {Smee}, {Smith}, {Snedden}, {Stone}, {Stoughton},
  {Strauss}, {Stubbs}, {SubbaRao}, {Szalay}, {Szapudi}, {Szokoly}, {Thakar},
  {Tremonti}, {Tucker}, {Uomoto}, {Vanden Berk}, {Vogeley}, {Waddell}, {Wang},
  {Watanabe}, {Weinberg}, {Yanny}, \& {Yasuda}}]{sdss}
{York} D.~G. {et~al.}, 2000, \aj, 120, 1579

\bibitem[{{Zibetti}, {Charlot} \& {Rix}(2009){Zibetti}, {Charlot}, \&
  {Rix}}]{zibetti09}
{Zibetti} S., {Charlot} S., {Rix} H.-W., 2009, \mnras, 400, 1181

\end{thebibliography}

\onecolumn
\newcommand\phn{\phantom{N}}

\begin{landscape}
\begin{table*}
\small
\caption{\label{t_det}\hi\ Properties of \hi\ Detections.}
\begin{tabular}{ccccccccccccll}
\hline\hline
      &                 & $T_{\rm on}$ & $\Delta v$ &     & \whi  & \whi$^c$&  $F$      &  {\it rms} & &Log \Mhi  &   &  & \\
AGC  & SDSS ID  & (min)       &  (\kms)    & $z$ &  (\kms)& (\kms) & (Jy \kms) & (mJy)& S/N  & (\Msun) & Log \Mhi/\Mst & Q & Project ID\\
(1)  & (2)  & (3)  & (4)  & (5)  & (6)  & (7)  & (8)  & (9)  &  (10) & (11) & (12) & (13) & (14)\\
\hline
101750 & J003610.70+142246.4 & 156 &  28 & 0.190872 & 543$\pm$  21 & 444 &  0.19$\pm$  0.03 &  0.12 &   8.2 & 10.51 & $-$0.47 &  1 &  a1803f	  \\
101710 & J003840.25+153912.9 &  88 &  27 & 0.176535 & 598$\pm$   4 & 497 &  0.16$\pm$  0.05 &  0.16 &   4.5 & 10.37 & $-$0.72 &  2 &  1803f	  \\
122040 & J020849.45+130442.5 &  56 &  27 & 0.168210 & 332$\pm$  23 & 273 &  0.34$\pm$  0.04 &  0.17 &  15.0 & 10.65 & \phn 0.28 &  1 &  a1803f	  \\
181518 & J082522.13+325953.6 & 112 &  27 & 0.171022 & 515$\pm$  23 & 428 &  0.26$\pm$  0.05 &  0.17 &   8.0 & 10.55 & $-$0.36 &  1 &  a1803	  \\
181593 & J084800.91+061837.2 & 176 &  29 & 0.220272 & 256$\pm$   9 & 198 &  0.13$\pm$  0.04 &  0.18 &   6.0 & 10.47 & $-$0.84 &  1 &  a2008,a2270 \\
181559 & J085733.51+034456.8 & 132 &  28 & 0.189094 & 546$\pm$  13 & 447 &  0.25$\pm$  0.05 &  0.17 &   7.4 & 10.63 & $-$0.41 &  1 &  a1803	  \\
198310 & J090315.37+073332.4 & 230 &  29 & 0.217517 & 481$\pm$   3 & 383 &  0.15$\pm$  0.03 &  0.12 &   7.2 & 10.53 & $-$0.56 &  1 &  a2270	  \\
191826 & J091954.42+085344.0 & 116 &  28 & 0.189905 & 770$\pm$  21 & 636 &  0.42$\pm$  0.06 &  0.17 &   8.7 & 10.85 & $-$0.34 &  1 &  a2008	  \\
191728 & J091957.00+013851.6 &  96 &  27 & 0.176425 & 219$\pm$   6 & 174 &  0.18$\pm$  0.03 &  0.18 &   9.3 & 10.42 & $-$0.47 &  1 &  a1803	  \\
191787 & J094921.53+044239.7 & 104 &  28 & 0.190145 & 601$\pm$   5 & 493 &  0.26$\pm$  0.06 &  0.18 &   6.2 & 10.64 & $-$0.33 &  1 &  a1803	  \\
191838 & J095923.26+065109.1 &  72 &  38 & 0.189998 & 538$\pm$  13 & 436 &  0.19$\pm$  0.06 &  0.16 &   5.0 & 10.51 & $-$0.52 &  3 &  a2008	  \\
202125 & J102848.86+042331.9 & 112 &  28 & 0.207010 & 480$\pm$  17 & 386 &  0.10$\pm$  0.04 &  0.16 &   3.4 & 10.29 & $-$0.81 &  3 &  a1803	  \\
208750 & J105138.96+061751.7 &  90 &  41 & 0.223601 & 466$\pm$  79 & 364 &  0.15$\pm$  0.06 &  0.17 &   4.3 & 10.56 & $-$0.69 &  2 &  a2428	  \\
208751 & J105230.22+082324.3 & 110 &  28 & 0.187710 & 492$\pm$  23 & 403 &  0.19$\pm$  0.04 &  0.14 &   7.6 & 10.51 & $-$0.42 &  1 &  a2428	  \\
212966 & J110730.34+072350.2 & 124 &  36 & 0.206446 & 494$\pm$  16 & 395 &  0.13$\pm$  0.04 &  0.14 &   4.4 & 10.41 & $-$0.69 &  2 &  a2008	  \\
212967 & J111353.47+094959.7 & 150 &  34 & 0.247755 & 414$\pm$   5 & 318 &  0.15$\pm$  0.04 &  0.16 &   5.4 & 10.63 & \phn 0.01 &  3 &  a2270	  \\
212941 & J111645.15+054210.0 & 180 &  29 & 0.223625 & 383$\pm$  39 & 301 &  0.18$\pm$  0.04 &  0.17 &   7.0 & 10.63 & $-$0.36 &  1 &  a1803	  \\
212887 & J114634.33+053442.0 & 176 &  28 & 0.195732 & 772$\pm$  23 & 634 &  0.29$\pm$  0.04 &  0.11 &   9.0 & 10.72 & $-$0.25 &  1 &  a1803	  \\
212871 & J120052.83+033754.9 & 100 &  28 & 0.190208 & 742$\pm$   7 & 612 &  0.30$\pm$  0.06 &  0.17 &   6.2 & 10.71 & $-$0.56 &  1 &  a1803	  \\
224321 & J120948.14+100822.5 &  80 &  28 & 0.187520 & 417$\pm$  14 & 339 &  0.26$\pm$  0.05 &  0.19 &   8.9 & 10.64 & $-$0.29 &  1 &  a2008	  \\
228389 & J122036.32+081904.2 &  75 &  29 & 0.221760 & 377$\pm$  10 & 297 &  0.17$\pm$  0.05 &  0.18 &   6.3 & 10.60 & $-$0.44 &  1 &  a2428	  \\
239110 & J130434.51+271334.9 &  75 &  29 & 0.220002 & 533$\pm$   3 & 425 &  0.30$\pm$  0.05 &  0.18 &   8.3 & 10.84 & $-$0.26 &  1 &  a2428	  \\
232041 & J132325.27+043707.6 & 203 &  27 & 0.186225 & 591$\pm$  29 & 487 &  0.19$\pm$  0.04 &  0.12 &   7.5 & 10.50 & $-$0.55 &  1 &  a1803,a2270 \\
232046 & J133327.34+042244.8 & 144 &  31 & 0.188360 & 818$\pm$   5 & 675 &  0.24$\pm$  0.05 &  0.14 &   5.1 & 10.60 & $-$0.38 &  2 &  a1803,a2008 \\
232055 & J134211.36+053211.5 &  92 &  28 & 0.202987 & 464$\pm$  10 & 374 &  0.26$\pm$  0.05 &  0.18 &   8.4 & 10.70 & $-$0.12 &  1 &  a2008	  \\
232127 & J135949.19+052731.8 & 140 &  37 & 0.228068 & 604$\pm$  17 & 477 &  0.22$\pm$  0.05 &  0.14 &   6.1 & 10.74 & $-$0.50 &  1 &  a2270	  \\
242091 & J140522.72+052814.6 & 156 &  28 & 0.195302 & 502$\pm$   4 & 408 &  0.25$\pm$  0.04 &  0.13 &  10.3 & 10.65 & $-$0.38 &  1 &  a1803	  \\
242073 & J140856.04+051616.3 & 140 &  28 & 0.190625 & 422$\pm$   2 & 343 &  0.15$\pm$  0.03 &  0.13 &   7.4 & 10.42 & $-$0.37 &  1 &  a2270	  \\
249558 & J142151.16+100623.6 &  85 &  28 & 0.193841 & 383$\pm$   6 & 309 &  0.15$\pm$  0.03 &  0.14 &   7.2 & 10.42 & $-$0.87 &  1 &  a2428	  \\
242147 & J142735.69+033434.2 & 176 &  30 & 0.245370 & 499$\pm$  11 & 389 &  0.28$\pm$  0.05 &  0.18 &   8.0 & 10.90 & $-$0.36 &  1 &  a1803	  \\
249559 & J144518.88+025012.3 &  95 &  28 & 0.190355 & 331$\pm$  73 & 266 &  0.18$\pm$  0.04 &  0.18 &   7.3 & 10.47 & $-$0.70 &  1 &  a2428	  \\
249560 & J144823.96+125551.9 & 164 &  35 & 0.195415 & 771$\pm$   9 & 630 &  0.19$\pm$  0.04 &  0.12 &   5.0 & 10.53 & $-$0.55 &  3 &  a2428	  \\
252580 & J151337.28+041921.1 &  84 &  27 & 0.174961 & 478$\pm$   8 & 396 &  0.32$\pm$  0.06 &  0.21 &   8.4 & 10.66 & $-$0.12 &  1 &  a1803	  \\
252654 & J153240.56+323428.8 &  92 &  32 & 0.201136 & 792$\pm$   4 & 646 &  0.26$\pm$  0.06 &  0.16 &   5.0 & 10.69 & $-$0.35 &  2 &  a2008	  \\
252641 & J153400.67+001133.2 & 116 &  28 & 0.203574 & 591$\pm$  86 & 479 &  0.31$\pm$  0.09 &  0.29 &   4.8 & 10.78 & $-$0.24 &  3 &  a1803	  \\
262026 & J160433.51+310900.0 & 112 &  41 & 0.223168 & 642$\pm$   2 & 509 &  0.13$\pm$  0.05 &  0.14 &   3.2 & 10.49 & $-$0.85 &  3 &  a2008,a2270 \\
262029 & J160938.00+312958.5 & 136 &  29 & 0.218648 & 642$\pm$  56 & 515 &  0.34$\pm$  0.05 &  0.15 &   9.0 & 10.89 & $-$0.41 &  1 &  a2008	  \\
262033 & J162515.41+280530.3 & 260 &  28 & 0.187113 & 643$\pm$  43 & 530 &  0.24$\pm$  0.03 &  0.11 &   9.4 & 10.59 & $-$0.49 &  1 &  a2270	  \\
262015 & J165940.12+344307.8 &  96 &  28 & 0.198331 & 504$\pm$  16 & 409 &  0.42$\pm$  0.05 &  0.19 &  12.0 & 10.89 & \phn 0.06 &  1 &  a1803	  \\
\hline
\end{tabular}
\end{table*}
\end{landscape}

\twocolumn

\onecolumn
\begin{landscape}
\begin{table*}
\small
\centering
\caption{\label{t_sdss}SDSS and UV Parameters.}
\begin{tabular}{ccccccccccccccr}
\hline\hline
   &  &   & Log \Mst & $R_{50}$ & $R_{90}$ & $R_{90}$ & Log \must  & ext$_r$ & r & (b/a)$_r$ & incl & SFR &\nuvr &  \\
AGC  & SDSS ID &  $z_{\rm SDSS}$ & (\Msun) & (\arcsec)&(\arcsec) & (kpc) & (\Msun~kpc$^{-2}$)& (mag)& (mag) &   & (deg) & (\Msun~yr$^{-1}$) & (mag)&  GALEX \\
(1)  & (2)  & (3)  & (4)  & (5) & (6) & (7) & (8)  & (9)  &  (10) & (11) & (12) & (13) & (14) & (15) \\
\hline
101750 & J003610.70+142246.4 & 0.1909 & 10.98 &  2.42 &  5.17 &  23.3 &  8.11 &  0.18 & 17.52 &   0.463 &   65 &   6.4 &  3.11 &  MIS \\
101710 & J003840.25+153912.9 & 0.1769 & 11.09 &  2.84 &  7.07 &  29.3 &  8.15 &  0.16 & 17.59 &   0.390 &   70 &   2.6 &  3.34 &  MIS \\
122040 & J020849.45+130442.5 & 0.1682 & 10.37 &  1.72 &  4.48 &  17.6 &  7.91 &  0.25 & 17.63 &   0.363 &   72 &   8.0 &  1.91 &  MIS \\
181518 & J082522.13+325953.6 & 0.1709 & 10.91 &  2.50 &  5.37 &  21.4 &  8.11 &  0.12 & 17.49 &   0.429 &   67 &  10.4 &  2.43 &  MIS \\
181593 & J084800.91+061837.2 & 0.2201 & 11.31 &  1.77 &  3.97 &  21.0 &  8.57 &  0.14 & 17.51 &   0.603 &   55 &  35.4 &   ... &  ... \\
181559 & J085733.51+034456.8 & 0.1894 & 11.04 &  2.59 &  5.74 &  25.7 &  8.12 &  0.13 & 17.50 &   0.516 &   61 &  13.7 &  2.66 &  MIS \\
198310 & J090315.37+073332.4 & 0.2174 & 11.09 &  1.94 &  4.41 &  23.0 &  8.28 &  0.15 & 18.32 &   0.263 &   80 &  27.7 &  3.09 &  MIS \\
191826 & J091954.42+085344.0 & 0.1899 & 11.19 &  3.16 &  7.03 &  31.5 &  8.09 &  0.14 & 17.43 &   0.280 &   79 &  13.1 &  2.76 &  AIS \\
191728 & J091957.00+013851.6 & 0.1763 & 10.89 &  2.03 &  4.61 &  19.0 &  8.25 &  0.08 & 17.24 &   0.679 &   49 &  14.7 &  1.80 &  MIS \\
191787 & J094921.53+044239.7 & 0.1903 & 10.97 &  2.62 &  5.57 &  25.0 &  8.03 &  0.11 & 17.66 &   0.475 &   64 &   7.5 &  2.66 &  MIS \\
191838 & J095923.26+065109.1 & 0.1902 & 11.03 &  1.83 &  4.34 &  19.5 &  8.40 &  0.11 & 17.69 &   0.455 &   65 &  11.1 &  2.81 &  AIS \\
202125 & J102848.86+042331.9 & 0.2069 & 11.10 &  2.84 &  5.52 &  27.2 &  8.01 &  0.09 & 17.74 &   0.341 &   74 &   7.9 &  2.95 &  MIS \\
208750 & J105138.96+061751.7 & 0.2239 & 11.25 &  3.52 &  7.02 &  37.9 &  7.90 &  0.09 & 17.31 &   0.811 &   37 &  25.1 &  2.67 &  MIS \\
208751 & J105230.22+082324.3 & 0.1880 & 10.93 &  2.47 &  5.01 &  22.2 &  8.06 &  0.07 & 17.57 &   0.461 &   65 &   4.7 &  2.38 &  MIS \\
212966 & J110730.34+072350.2 & 0.2066 & 11.10 &  1.36 &  3.21 &  15.8 &  8.65 &  0.09 & 17.78 &   0.607 &   54 &  16.4 &  3.05 &  AIS \\
212967 & J111353.47+094959.7 & 0.2482 & 10.62 &  1.50 &  3.47 &  21.0 &  7.90 &  0.06 & 18.40 &   0.683 &   48 &  12.7 &  2.02 &  AIS \\
212941 & J111645.15+054210.0 & 0.2239 & 10.99 &  2.73 &  4.91 &  26.5 &  7.86 &  0.21 & 17.53 &   0.589 &   56 &  11.9 &  1.85 &  AIS \\
212887 & J114634.33+053442.0 & 0.1961 & 10.97 &  2.03 &  4.48 &  20.8 &  8.22 &  0.05 & 17.78 &   0.360 &   72 &   9.7 &   ... &  ... \\
212871 & J120052.83+033754.9 & 0.1904 & 11.27 &  2.93 &  6.75 &  30.4 &  8.23 &  0.07 & 17.23 &   0.427 &   67 &  24.7 &   ... &  ... \\
224321 & J120948.14+100822.5 & 0.1879 & 10.93 &  2.93 &  6.40 &  28.4 &  7.90 &  0.07 & 17.27 &   0.698 &   47 &  11.5 &  2.58 &  AIS \\
228389 & J122036.32+081904.2 & 0.2218 & 11.04 &  2.42 &  5.36 &  28.6 &  8.02 &  0.06 & 17.70 &   0.636 &   52 &   4.2 &  1.99 &  MIS \\
239110 & J130434.51+271334.9 & 0.2201 & 11.10 &  4.48 &  8.41 &  44.5 &  7.55 &  0.03 & 17.33 &   0.630 &   52 &   4.9 &  2.28 &  MIS \\
232041 & J132325.27+043707.6 & 0.1861 & 11.05$^*$ &  1.98 &  4.59 &  20.1 &  9.69 &  0.08 & 17.53 &   0.511 &   61 &  20.7 &   ... &  ... \\
232046 & J133327.34+042244.8 & 0.1888 & 10.98 &  2.43 &  5.68 &  25.3 &  8.11 &  0.08 & 17.60 &   0.380 &   71 &   6.0 &  2.78 &  AIS \\
232055 & J134211.36+053211.5 & 0.2031 & 10.82 &  2.36 &  5.45 &  26.3 &  7.90 &  0.08 & 17.75 &   0.540 &   59 &   9.1 &  2.31 &  AIS \\
232127 & J135949.19+052731.8 & 0.2281 & 11.24 &  2.36 &  5.09 &  28.0 &  8.21 &  0.08 & 17.70 &   0.607 &   54 &  17.9 &  3.06 &  MIS \\
242091 & J140522.72+052814.6 & 0.1954 & 11.03 &  3.12 &  6.42 &  29.7 &  7.91 &  0.07 & 17.23 &   0.458 &   65 &  17.5 &  2.63 &  MIS \\
242073 & J140856.04+051616.3 & 0.1906 & 10.79 &  2.11 &  4.61 &  20.8 &  8.03 &  0.07 & 17.66 &   0.576 &   57 &  12.7 &  2.37 &  MIS \\
249558 & J142151.16+100623.6 & 0.1938 & 11.29 &  3.33 &  6.54 &  30.0 &  8.13 &  0.08 & 17.00 &   0.790 &   39 &  16.8 &  2.62 &  MIS \\
242147 & J142735.69+033434.2 & 0.2455 & 11.26 &  2.49 &  5.87 &  35.1 &  8.11 &  0.09 & 17.54 &   0.629 &   53 &  26.6 &  2.49 &  MIS \\
249559 & J144518.88+025012.3 & 0.1906 & 11.17 &  2.62 &  6.21 &  28.0 &  8.23 &  0.11 & 17.28 &   0.773 &   40 &  15.6 &  3.01 &  MIS \\
249560 & J144823.96+125551.9 & 0.1958 & 11.08 &  3.08 &  6.33 &  29.4 &  7.97 &  0.07 & 17.41 &   0.717 &   45 &   9.8 &  2.70 &  MIS \\
252580 & J151337.28+041921.1 & 0.1754 & 10.78 &  3.60 &  7.94 &  32.6 &  7.64 &  0.13 & 17.57 &   0.441 &   66 &   6.1 &  2.38 &  MIS \\
252654 & J153240.56+323428.8 & 0.2001 & 11.04 &  2.44 &  5.33 &  25.3 &  8.12 &  0.06 & 17.66 &   0.365 &   72 &  16.3 &   ... &  ... \\
252641 & J153400.67+001133.2 & 0.2036 & 11.02 &  2.24 &  4.84 &  23.5 &  8.15 &  0.21 & 17.49 &   0.547 &   59 &   6.2 &  2.07 &  MIS \\
262026 & J160433.51+310900.0 & 0.2228 & 11.34 &  3.18 &  6.82 &  36.6 &  8.07 &  0.08 & 17.16 &   0.507 &   62 &  26.3 &  2.75 &  AIS \\
262029 & J160938.00+312958.5 & 0.2188 & 11.30 &  2.87 &  6.44 &  33.8 &  8.15 &  0.09 & 17.20 &   0.608 &   54 &  20.5 &  2.84 &  AIS \\
262033 & J162515.41+280530.3 & 0.1870 & 11.08 &  3.26 &  6.82 &  30.1 &  7.97 &  0.19 & 17.66 &   0.247 &   82 &   5.1 &   ... &  ... \\
262015 & J165940.12+344307.8 & 0.1982 & 10.83 &  2.69 &  5.59 &  26.3 &  7.83 &  0.06 & 17.54 &   0.591 &   55 &   7.7 &  2.15 &  MIS \\
\hline
\end{tabular}
\begin{flushleft}
$^{*}$SDSS value was 12.37 (see text).
\end{flushleft}
\end{table*}
\end{landscape}

\twocolumn


\appendix

\section{SDSS images and Arecibo HI spectra}\label{s_spectra}

We present here SDSS postage stamp images and Arecibo \hi-line spectra
for the 39 \highz\ galaxies discussed in this work.
These are organized as follows: Fig.~\ref{det} shows the
28 \hi\ detections with quality flag $\rm Q=1$ in Table~\ref{t_det}, 
and Fig.~\ref{dcode2} and \ref{dcode3} show 5 galaxies with $\rm Q=2$ and 
6 marginal detections with $\rm Q=3$, respectively.
The objects in each of these figures are ordered by 
increasing AGC number (indicated on the top right corner of the SDSS image).
The SDSS images show a 1 arcmin square field, \ie\ only the central
part of the region sampled by the Arecibo beam (the half
power full width of the beam is \about 4\arcmin\ at the
frequencies of these observations). 
The \hi\ spectra are always displayed over a velocity
interval corresponding to the full 12.5 MHz bandwidth adopted for our
observations. The \hi-line profiles are calibrated, smoothed 
(to a velocity resolution between 27 and 41 \kms, as listed in Table~\ref{t_det}), and
baseline-subtracted. A red, dotted line indicates the heliocentric
velocity corresponding to the optical redshift from SDSS. 
The shaded area and two vertical
dashes show the part of the profile that was integrated to
measure the \hi\ flux and the peaks used for width measurement,
respectively.

\begin{figure*}
\includegraphics[width=17.5cm]{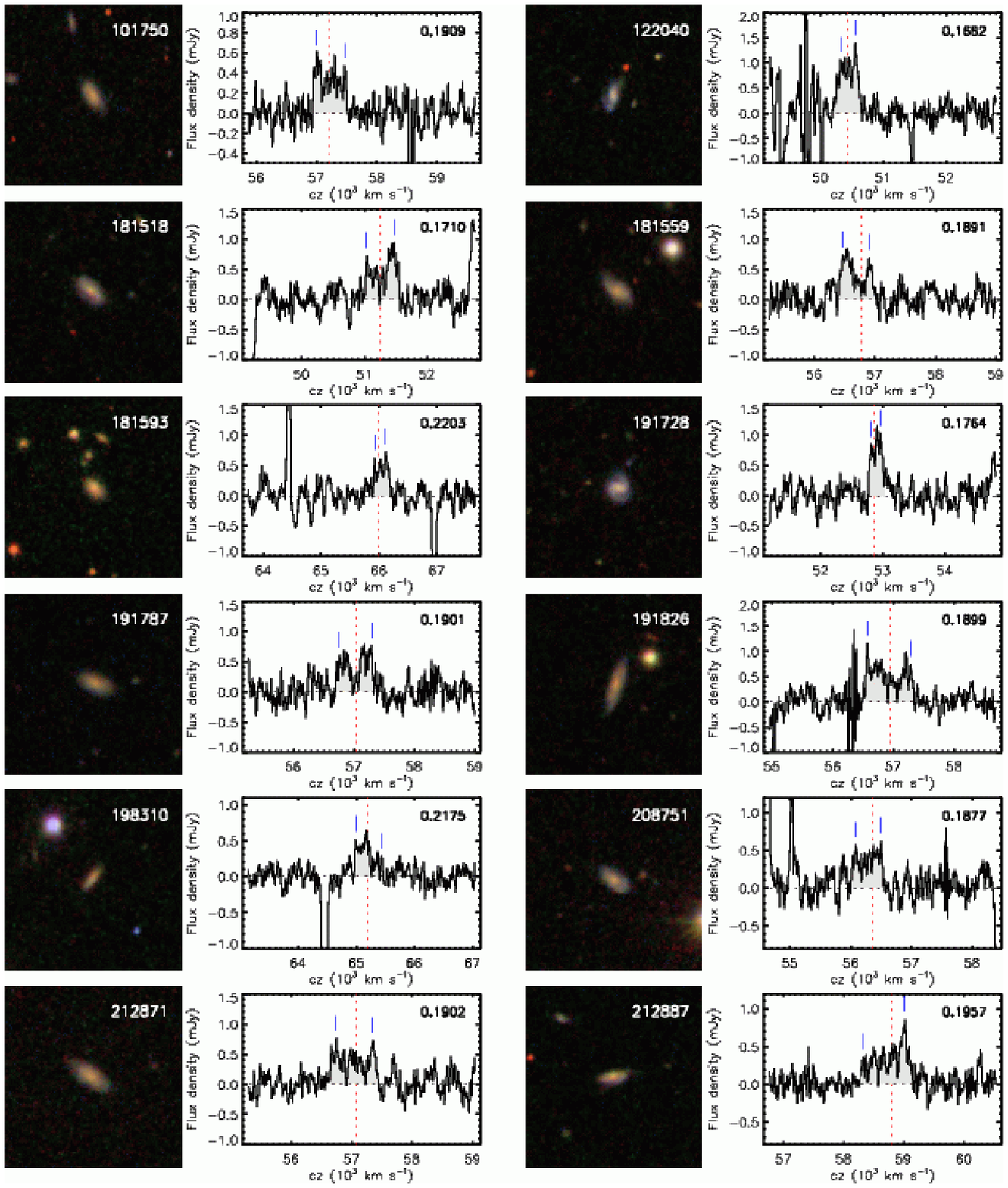}
\caption{SDSS postage stamp images (1 arcmin square) and
\hi-line profiles of the best quality detections (code 1) presented in this work,
ordered by increasing AGC number (indicated on each image). The \hi\ spectra are
calibrated, smoothed and baseline-subtracted. A dotted line and two
dashes indicate the heliocentric velocity corresponding to the SDSS
redshift and the two peaks used for width measurement,
respectively. The redshift measured from the \hi\ profile is indicated
on the top right corner of the spectrum.}
\label{det}
\end{figure*}

\setcounter{figure}{0}
\begin{figure*}
\includegraphics[width=17.5cm]{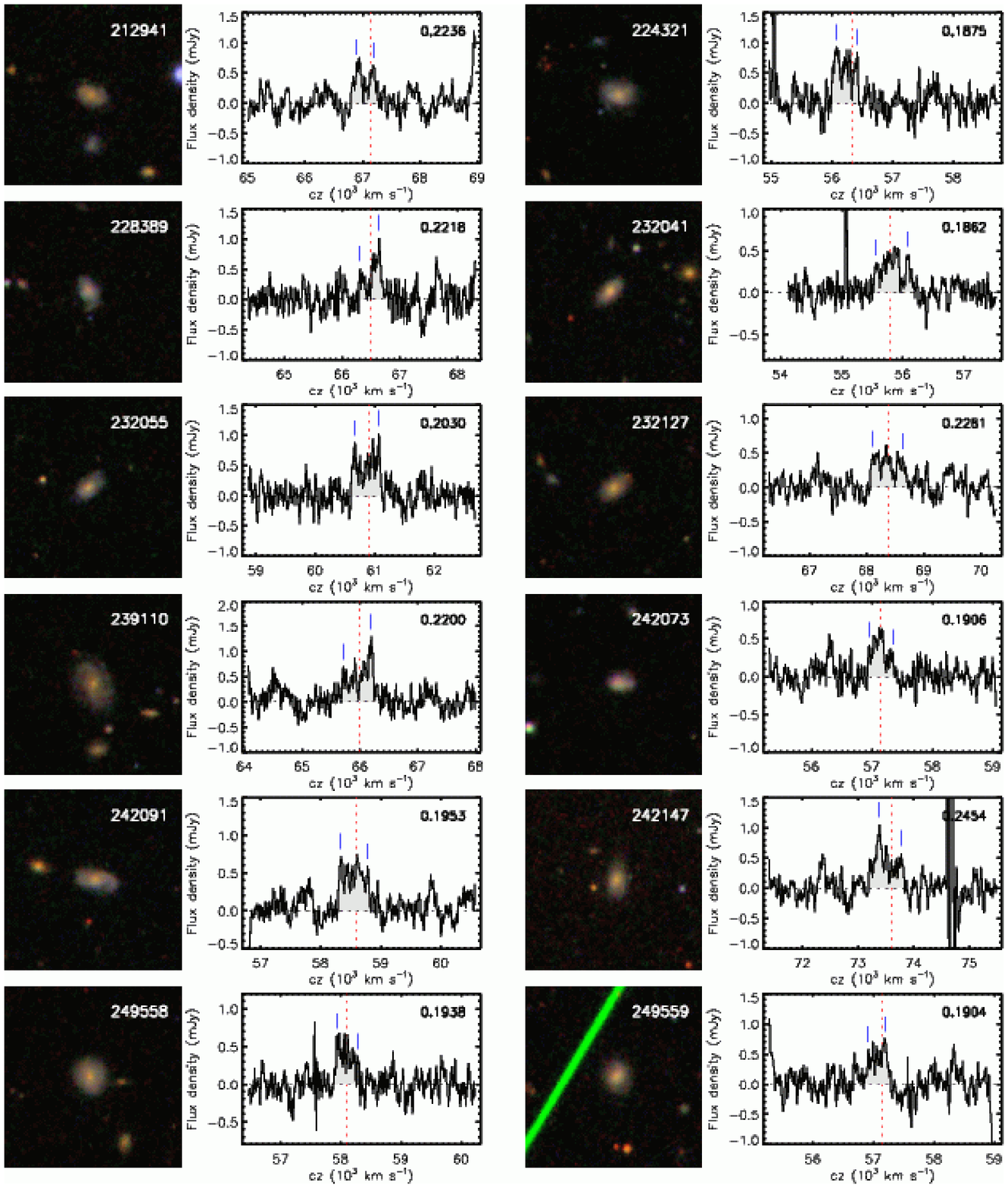}
\caption{\it continued}
\end{figure*}

\setcounter{figure}{0}
\begin{figure*}
\includegraphics[width=17.5cm]{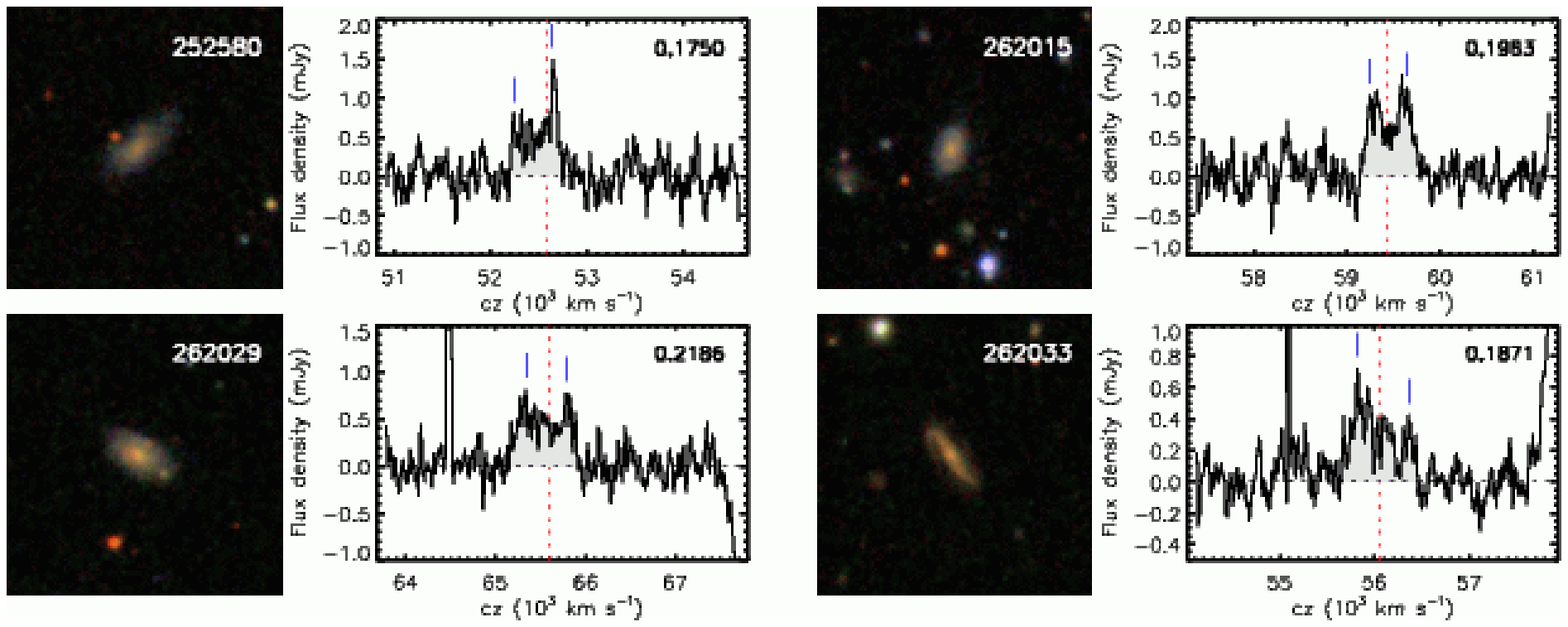}
\caption{\it continued}
\end{figure*}

\setcounter{figure}{1}
\begin{figure*}
\includegraphics[width=17.5cm]{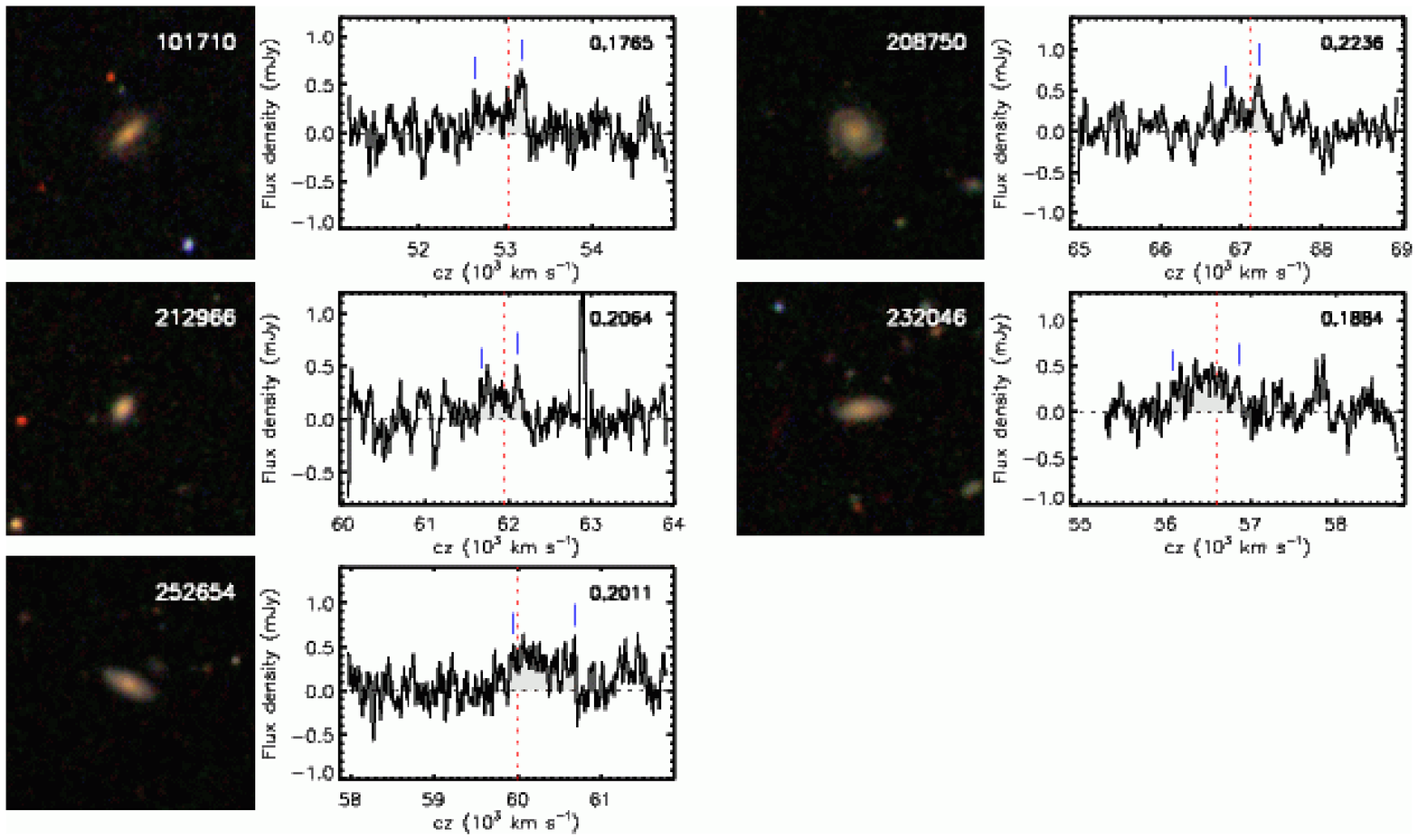}
\caption{Same as Fig.~\ref{det} for lower signal-to-noise detections (code 2).}
\label{dcode2}
\end{figure*}

\setcounter{figure}{2}
\begin{figure*}
\includegraphics[width=17.5cm]{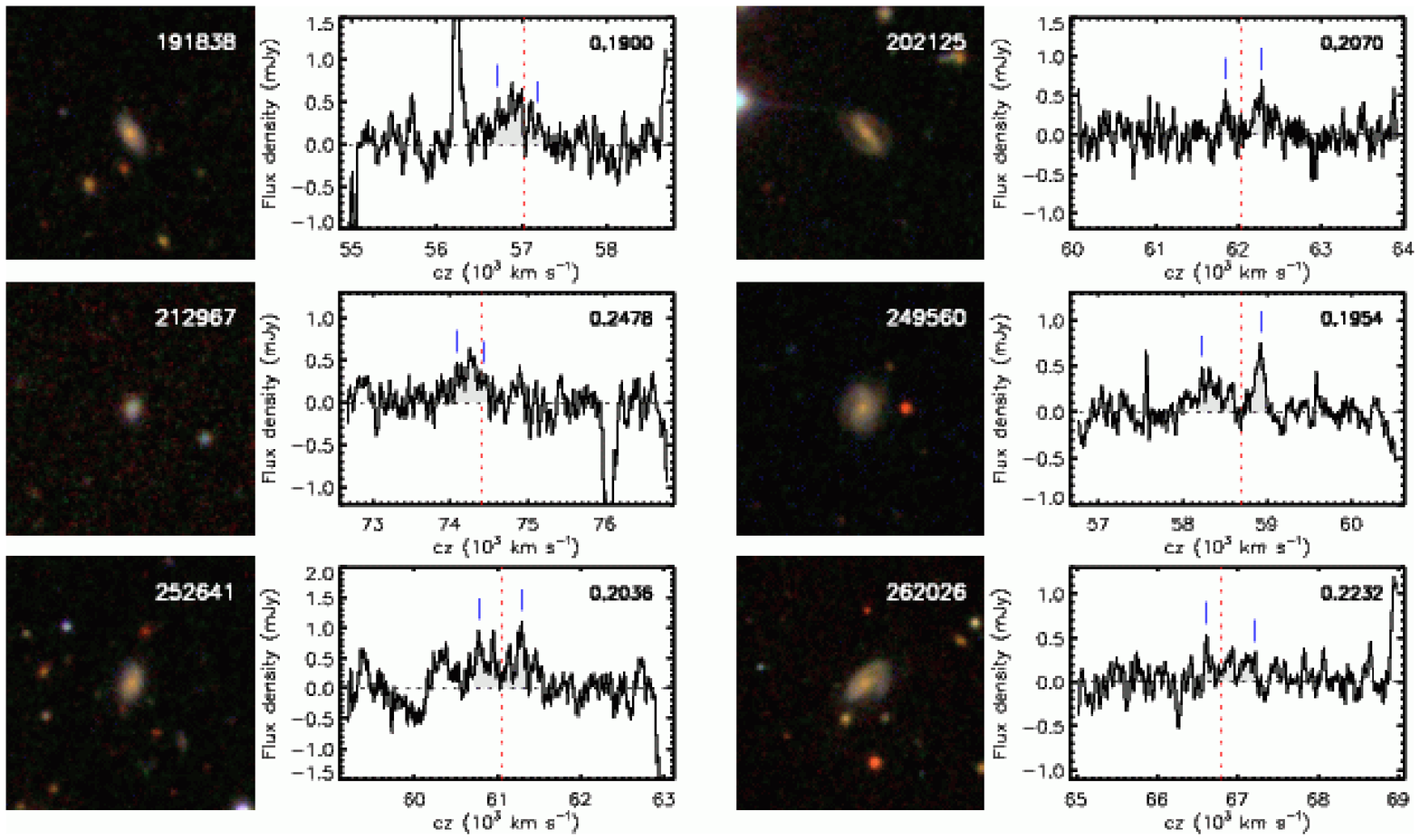}
\caption{Same as Fig.~\ref{det} for marginal detections (code 3).}
\label{dcode3}
\end{figure*}


\end{document}